\documentstyle[12pt]{article}

\def\hybrid{\topmargin 0pt	\oddsidemargin 0pt
	\headheight 0pt	\headsep 0pt
	\textheight 9in		
	\textwidth 6.25in	
	\marginparwidth .875in
	\parskip 5pt plus 1pt	\jot = 1.5ex}

\hybrid

\def\baselinestretch{1.2}

\catcode`\@=11

\def\marginnote#1{}
\newcount\hour
\newcount\minute
\newtoks\amorpm
\hour=\time\divide\hour by60
\minute=\time{\multiply\hour by60 \global\advance\minute by-\hour}
\edef\standardtime{{\ifnum\hour<12 \global\amorpm={am}%
	\else\global\amorpm={pm}\advance\hour by-12 \fi
	\ifnum\hour=0 \hour=12 \fi
	\number\hour:\ifnum\minute<10 0\fi\number\minute\the\amorpm}}
\edef\militarytime{\number\hour:\ifnum\minute<10 0\fi\number\minute}

\def\draftlabel#1{{\@bsphack\if@filesw {\let\thepage\relax
   \xdef\@gtempa{\write\@auxout{\string
      \newlabel{#1}{{\@currentlabel}{\thepage}}}}}\@gtempa
   \if@nobreak \ifvmode\nobreak\fi\fi\fi\@esphack}
	\gdef\@eqnlabel{#1}}
\def\@eqnlabel{}
\def\@vacuum{}
\def\draftmarginnote#1{\marginpar{\raggedright\scriptsize\tt#1}}

\def\draft{\oddsidemargin -.5truein
	\def\@oddfoot{\sl preliminary draft \hfil
	\rm\thepage\hfil\sl\today\quad\militarytime}
	\let\@evenfoot\@oddfoot	\overfullrule 3pt
	\let\label=\draftlabel
	\let\marginnote=\draftmarginnote
   \def\@eqnnum{(\theequation)\rlap{\kern\marginparsep\tt\@eqnlabel}%
\global\let\@eqnlabel\@vacuum}  }

\def\preprint{\twocolumn\sloppy\flushbottom\parindent 2em
	\leftmargini 2em\leftmarginv .5em\leftmarginvi .5em
	\oddsidemargin -.5in	\evensidemargin -.5in
	\columnsep .4in	\footheight 0pt
	\textwidth 10in	\topmargin  -.4in
	\headheight 12pt \topskip .4in
	\textheight 7.1in \footskip 0pt
	\def\@oddhead{\thepage\hfil\addtocounter{page}{1}\thepage}
	\let\@evenhead\@oddhead	\def\@oddfoot{}	\def\@evenfoot{} }

\def\numberbysection{\@addtoreset{equation}{section}
	\def\theequation{\thesection.\arabic{equation}}}

\def\underline#1{\relax\ifmmode\@@underline#1\else
	$\@@underline{\hbox{#1}}$\relax\fi}

\def\titlepage{\@restonecolfalse\if@twocolumn\@restonecoltrue\onecolumn
     \else \newpage \fi \thispagestyle{empty}\c@page\z@
	\def\thefootnote{\fnsymbol{footnote}} }

\def\endtitlepage{\if@restonecol\twocolumn \else \newpage \fi
	\def\thefootnote{\arabic{footnote}}
	\setcounter{footnote}{0}}  

\catcode`@=12
\relax

\def\figcap{\section*{Figure Captions\markboth
	{FIGURECAPTIONS}{FIGURECAPTIONS}}\list
	{Figure \arabic{enumi}:\hfill}{\settowidth\labelwidth{Figure 999:}
	\leftmargin\labelwidth
	\advance\leftmargin\labelsep\usecounter{enumi}}}
 \relax
\def\tablecap{\section*{Table Captions\markboth
	{TABLECAPTIONS}{TABLECAPTIONS}}\list
	{Table \arabic{enumi}:\hfill}{\settowidth\labelwidth{Table 999:}
	\leftmargin\labelwidth
	\advance\leftmargin\labelsep\usecounter{enumi}}}
 \relax
\def\reflist{\section*{References\markboth
	{REFLIST}{REFLIST}}\list
	{[\arabic{enumi}]\hfill}{\settowidth\labelwidth{[999]}
	\leftmargin\labelwidth
	\advance\leftmargin\labelsep\usecounter{enumi}}}
 \relax
\makeatletter
\newcounter{pubctr}
\def\publist{\@ifnextchar[{\@publist}{\@@publist}}
\def\@publist[#1]{\list
	{[\arabic{pubctr}]\hfill}{\settowidth\labelwidth{[999]}
	\leftmargin\labelwidth
	\advance\leftmargin\labelsep
	\@nmbrlisttrue\def\@listctr{pubctr}
	\setcounter{pubctr}{#1}\addtocounter{pubctr}{-1}}}
\def\@@publist{\list
	{[\arabic{pubctr}]\hfill}{\settowidth\labelwidth{[999]}
	\leftmargin\labelwidth
	\advance\leftmargin\labelsep
	\@nmbrlisttrue\def\@listctr{pubctr}}}
 \relax
\makeatother
\newskip\humongous \humongous=0pt plus 1000pt minus 1000pt
\def\caja{\mathsurround=0pt}

\newif\ifdtup
\def\panorama{\global\dtuptrue \openup1\jot \caja
	\everycr{\noalign{\ifdtup \global\dtupfalse
	\vskip-\lineskiplimit \vskip\normallineskiplimit
	\else \penalty\interdisplaylinepenalty \fi}}}
\def\eqalignno#1{\panorama \tabskip=\humongous
	\halign to\displaywidth{\hfil$\displaystyle{##}$
	\tabskip=0pt&$\displaystyle{{}##}$\hfil
	\tabskip=\humongous&\llap{$##$}\tabskip=0pt
	\crcr#1\crcr}}
\relax
\font\bfi=cmmib10

\def\diagarrowr{\nwarrow \hskip-10.5pt \searrow }
\def\diagarrowl{\swarrow \hskip-10.5pt \nearrow }
\def\mapdiagr#1{\diagarrowr {\scriptstyle#1} }
\def\mapdiagl#1{\diagarrowl {\scriptstyle#1} }

\def\a{\alpha}
\def\hh{{\hat h}}
\def\au{Aut\,g}
\def\b{\beta}
\def\d{\delta}
\def\e{\epsilon}
\def\m{\eta}

\def\s{\sigma}
\def\hh{\hat{h}}

\def\xx{\hbox{ }^*_*}

\def\o{\omega}

\def\bp{\vec p}
\def\bq{\vec q}
\def\br{\vec r}

\def\u{\underline}
\def\mapright#1{\smash{\mathop{\longrightarrow}\limits^{#1}}}

\def\mapupdown#1{\Big\updownarrow\rlap{$\vcenter{\hbox{$\scriptstyle#1$}}$}}

\def\mapleft#1{\smash{\mathop{\longleftarrow}\limits^{#1}}}
\def\qed{\hbox{${\vcenter{\vbox{
    \hrule height 0.4pt\hbox{\vrule width 0.4pt height 6pt
    \kern5pt\vrule width 0.4pt}\hrule height 0.4pt}}}$}}

\def\N{{\rm I\!N}}

\def\Q{{\mathchoice
{\setbox0=\hbox{$\displaystyle\rm Q$}\hbox{\raise 0.15\ht0\hbox to0pt
{\kern0.4\wd0\vrule height0.8\ht0\hss}\box0}}
{\setbox0=\hbox{$\textstyle\rm Q$}\hbox{\raise 0.15\ht0\hbox to0pt
{\kern0.4\wd0\vrule height0.8\ht0\hss}\box0}}
{\setbox0=\hbox{$\scriptstyle\rm Q$}\hbox{\raise 0.15\ht0\hbox to0pt
{\kern0.4\wd0\vrule height0.7\ht0\hss}\box0}}
{\setbox0=\hbox{$\scriptscriptstyle\rm Q$}\hbox{\raise 0.15\ht0\hbox to0pt
{\kern0.4\wd0\vrule height0.7\ht0\hss}\box0}}}}
\def\C{{\mathchoice
{\setbox0=\hbox{$\displaystyle\rm C$}\hbox{\hbox to0pt
{\kern0.4\wd0\vrule height0.9\ht0\hss}\box0}}
{\setbox0=\hbox{$\textstyle\rm C$}\hbox{\hbox to0pt
{\kern0.4\wd0\vrule height0.9\ht0\hss}\box0}}
{\setbox0=\hbox{$\scriptstyle\rm C$}\hbox{\hbox to0pt
{\kern0.4\wd0\vrule height0.9\ht0\hss}\box0}}
{\setbox0=\hbox{$\scriptscriptstyle\rm C$}\hbox{\hbox to0pt
{\kern0.4\wd0\vrule height0.9\ht0\hss}\box0}}}}

\font\fivesans=cmss10 at 4.61pt
\font\sevensans=cmss10 at 6.81pt
\font\tensans=cmss10
\newfam\sansfam
\textfont\sansfam=\tensans\scriptfont\sansfam=\sevensans\scriptscriptfont
\sansfam=\fivesans
\def\sans{\fam\sansfam\tensans}
\def\Z{{\mathchoice
{\hbox{$\sans\textstyle Z\kern-0.4em Z$}}
{\hbox{$\sans\textstyle Z\kern-0.4em Z$}}
{\hbox{$\sans\scriptstyle Z\kern-0.3em Z$}}
{\hbox{$\sans\scriptscriptstyle Z\kern-0.2em Z$}}}}
\def\baselinestretch{1.2}
\def\thefootnote{\fnsymbol{footnote}}
\def\ref#1{$^{#1)}$}
\begin{document}
\begin{titlepage}
\begin{center}
September 1991   \hfill    UCB-PTH-91/47   \\
                 \hfill    LBL-31296\\
                 \hfill    BONN-HE-91-18 \\
{\large \bf The Lie ${\bf h}$-Invariant Conformal Field Theories \\
and the Lie ${\bf h}$-Invariant Graphs}
\footnote{This work was supported in part by the Director, Office of
Energy Research, Office of High Energy and Nuclear Physics, Division of
High Energy Physics of the U.S. Department of Energy under Contract
DE-AC03-76SF00098 and in part by the National Science Foundation under
grant PHY90-21139. The work of N. O. is supported in part by the Deutsche
Forschungsgemeinschaft.}
\vskip .2in

M. B. Halpern\footnote{email: HALPERN@LBL.bitnet,42070::HALPERN,
HALPERN@THEORM.LBL.gov},
E. B. Kiritsis\footnote{email: KIRITSIS@LBL.bitnet, 42070::KIRITSIS,
KIRITSIS@THEORM.LBL.gov}\footnote{Address after October 1, 1991: Lab. de
Physique Th\'eorique,
Ecole Normale Superieure, 24 rue Lhomond, F-75231, Paris, CEDEX 05, FRANCE.},
\vskip .1in
{\em  Department of Physics, University of California and\\
      Theoretical Physics Group, Lawrence Berkeley Laboratory\\
      Berkeley, CA 94720, USA}
\vskip .1in
and\\
\vskip .1in
N. A. Obers\footnote{email: 13581::OBERS, OBERS@DBNPIB5.bitnet}
\vskip .1in
{\em Physikalisches Institut der Universit\"at Bonn\\
Nussallee 12, D-5300 Bonn 1, GERMANY}
\end{center}
\vskip .1in

\begin{abstract}
We use the Virasoro master equation to study the space of Lie
$h$-invariant conformal field theories, which includes the standard
rational conformal field theories as a small subspace.
In a detailed example, we apply the general theory to characterize and study
the Lie $h$-invariant graphs, which classify the Lie $h$-invariant conformal
field theories of the diagonal ansatz on $SO(n)$.
The Lie characterization of these graphs is another aspect of the
recently observed Lie group-theoretic structure of graph theory.
\end{abstract}
\end{titlepage}
\renewcommand{\thepage}{\arabic{page}}
\setcounter{page}{1}
\renewcommand{\baselinestretch}{1.5}
\centerline{\large\bf Table of Contents}
\begin{tabbing}
\=1\= 0. \=Introduction\hfill\kill
\>\>1. \>Introduction\hfill\\
\>\>2. \>The Virasoro master equation\hfill\\
\>\>3. \>Group Invariant Ans\"atze\hfill\\
\>\>4. \>Lie $h$-Invariant Ans\"atze\hfill\\
\>\>5. \>(0,0) and (1,0) Operators in Lie $h$-Invariant CFTs\hfill\\
\>\>6. \>The Lie $h$-Invariant Conformal Multiplets\hfill\\
\>\>7. \>Lie Symmetry in Graph Theory\hfill\\
\>\>     \> 7.1 Strategy\hfill\\
\>\>     \> 7.2 A Review of $SO(n)_{diag}$\hfill\\
\>\>     \> 7.3 Lie $h$-Invariant Subans\"atze in $SO(n)_{diag}$\hfill\\
\>\>     \> 7.4 The Lie $h$-Invariant Graphs\hfill\\
\>\>     \> 7.5 On the Number of Lie $h$-Invariant Graphs\hfill\\
\>\>8. \>Graphical Identification of (0,0) and (1,0) Operators\hfill\\
\>\>9. The Lie $h$-Invariant Graph Multiplets\hfill\\
\>1\>0. Examples of Lie $h$-Invariant Graphs\hfill\\
\>\>   \> 10.1. Counting on Small Manifolds\hfill\\
\>\>   \>10.2. Self K$_{g/h}$-Complementary Graphs\hfill\\
\>1\>1. \>Exact Solutions on Lie $h$-Invariant Graphs\hfill\\
\>\>\> 11.1. Small Subans\"atze with New Level-Families\hfill\\
\>\>\> 11.2. An ($SO(2)\times SO(2)$)-Invariant Octet in $SO(6)_{diag}$\hfill\\
\>\>\> 11.3. An $SO(2)$-Invariant Quartet in $SO(6)_{diag}$\hfill\\
\>\>\> 11.4. The Self K$_{g/h}$-Conjugate Level-Families of
$SO(6)_{diag}$\hfill\\
\>Appendix A. Counting (0,0) and (1,0) Operators\hfill
\end{tabbing}
\renewcommand{\baselinestretch}{1.0}

\bigskip\noindent
{\large\bf 1.} {\large\bf Introduction}
\bigskip

Affine Lie algebra, or current algebra on $S^{1}$, was discovered independently
in mathematics \cite{1} and physics \cite{2}. The first representations
\cite{2} were obtained with world-sheet fermions \cite{2,3} in the
construction of current-algebraic spin and
internal symmetry on the string \cite{2}. Examples of
affine-Sugawara\footnote{The Sugawara-Sommerfield
model \cite{4} was in four dimensions on the algebra of fields.
The first affine-Sugawara constructions, on affine Lie algebra, were
given by Bardak\c ci and Halpern \cite{2,5} in 1971.}
constructions \cite{2,5} and coset constructions \cite{2,5} were also given in
the first string era, as well as the vertex operator construction of fermions
and $SU(n)_1$ from compactified spatial dimensions \cite{6,7}.
The generalization of these constructions \cite{8,9,10} and their application
to the heterotic string \cite{11} mark the beginning of the present era. See
\cite{12,13,14} for further historical remarks on affine-Virasoro
constructions.

The general Virasoro construction on the currents $J_a$ of affine $g$
\cite{15,16}
$$ T(L) = L^{ab}\xx J_a J_b \xx \eqno(1.1) $$
systematizes the algebraic approach used by Bardak\c ci and Halpern \cite{2,5}
to obtain the original affine-Sugawara and coset constructions.
The general construction\footnote{Beyond the addition of
$\partial J$ terms \cite{15}, more general Virasoro constructions may exist:
The most general Virasoro construction is  discussed abstractly in \cite{17}
and
the interacting bosonic constructions are discussed in \cite{17,18,19}.}
is summarized by the Virasoro master equation \cite{15,16} for the inverse
inertia tensor  $L^{ab}=L^{ba}$, which  contains the
affine-Sugawara nests \footnote{The affine-Sugawara nests \cite{20} include
the affine-Sugawara constructions \cite{2,5,9}, the coset constructions
\cite{2,5,10} and the nested coset constructions \cite{21}.}
and many new affine-Virasoro constructions $g^{\#}$ on the currents of
affine $g$.

In particular, broad classes of exact unitary solutions with irrational central
charge \cite{20} have recently been obtained on affine compact $g$.
The growing list
presently includes the unitary irrational constructions [20,22-28]
$$ ((\hbox{simply-laced}\,g_x)^q)^{\#}_M $$
$$ SU(3)^{\#}_{BASIC}=\left\{ \matrix{SU(3)^{\#}_M \cr
SU(3)_{D(1)}^{\#},\;\;SU(3)_{D(2)}^{\#},\;\;SU(3)_{D(3)}^{\#} \cr
SU(3)_{A(1)}^{\#},\;\;SU(3)_{A(2)}^{\#}} \right.  $$
$$ SO(n)^{\#}_{diag}=\left\{ \matrix{ SO(2n)^{\#}[d,4],\;\,n\geq 3 \cr
SO(5)^{\#}[d,6]_2 \cr
SO(2n+1)^{\#}[d,6]_{1,2},\;\,n \geq 3 \cr}\right. $$
$$ SU(5)^{\#}[m,2] $$
$$ SU(n)^{\#}[m(N=1),rs] $$
$$ SU(\Pi^s_i n_i)^{\#}[m(N=1);\{r\}\{t\}]\eqno(1.2)$$
\noindent
which are called conformal level-families because they are
defined for all levels of affine $g$.
As in general relativity, these level-families were obtained in various
ans\"atze and subans\"atze (BASIC$\supset$Dynkin$\supset$Maximal, diagonal,
metric, etc.) of the Virasoro master equation and its companion,
the superconformal master equation \cite{29}.

As an example, the value at level 5 of $SU(3)$ \cite{24}
$$ c\left( (SU(3)_5)^{\#}_{D(1)}\right) =2
\left(1-{1 \over \sqrt{61}}\right)
\simeq 1.7439 \eqno(1.3)$$
is the lowest unitary irrational central charge yet obtained.
The simplest exact unitary irrational level-families yet obtained are the
$rs$-superconformal set in (1.2) with central charge \cite{27}
$$ c(SU(n)^{\#}_x[m(N=1),rs])={6nx \over nx + 8\sin^2(rs \pi/n) } \eqno(1.4)$$
where $r, s \in \N$ and $x$ is the level of affine $SU(n)$.

More generally, the solution space of the master equation is immense, with a
very large number \cite{20}
$$N(g)=2^{{\fam=0 dim}g({\fam=0 dim}g-1)/2}\eqno(1.5)$$
of level-families expected generically on any $g$.
The level-families have generically irrational central charge
and it is a puzzle that the level-families that have been observed on compact
$g$, either exactly or by high-level expansion \cite{24}, are
generically unitary.

In order to classify their conformal field theories, the Virasoro master
equation and the superconformal master equation are generating generalized
graph theories on Lie $g$ [25-31], including conventional graph theory as a
special case on the orthogonal groups.
In this development, the graphs and generalized graphs not only classify large
sets of generically unitary and irrational conformal level-families, but an
underlying Lie group-theoretic structure is seen in each of the graph theories.
We mention in particular the Lie-algebraic form of the edge-adjacency matrix
and isomorphism groups of the graph theories, and we refer the interested
reader to Ref. \cite{28}, which axiomatizes the subject.
See also Ref. \cite{32} for a review of this and other developments in the
Virasoro and superconformal master equations.

In this paper, we study the $Lie$ $h$-$invariant$ $conformal$ $field$
$theories$, that
is, the solutions of the Virasoro master equation which possess a Lie
symmetry, with associated (0,0) and (1,0) operators.
Known examples in this class are the affine-Sugawara nests and the exact
$U(1)$-invariant level-families \cite{23,24}
$$SU(3)^{\#}_{D(1)}\;,\;SU(3)^{\#}_{D(2)}\;,\;SU(3)^{\#}_{A(1)}\;,\;SU(3)
^{\#}_{A(2)}$$
which are generically unitary with generically irrational central charge.
In fact, the set of Lie $h$-invariant conformal field
theories has generically irrational central charge, and is much larger than the
set of affine-Sugawara nests,
although Lie symmetry is not generic in the space of conformal field theories.

The space of Lie $h$-invariant conformal field theories is organized by
K-conjugation \cite{2,5,10,15} and Lie symmetry into sets of $Lie$
$h$-$invariant$ $conformal$
$multiplets$, which exhibit a large variety of generalized K-conjugations
through coset constructions and general affine-Sugawara nests.
We mention in particular the $self$ K$_{g/h}$-$conjugate$
$constructions$, whose central charges
$$c={1\over 2}c_{g/h}\eqno(1.6)$$
are half that of the corresponding $g/h$ coset constructions.
These constructions generalize the self K-conjugate constructions
\cite{25,26,33} which have half affine-Sugawara central charge.

As an example, we apply the general theory of Lie $h$-invariant conformal
constructions to the original graph-theory ansatz $SO(n)_{diag}$ \cite{25},
whose generically unitary and irrational level-families are classified by the
conventional graphs of order $n$.
The Lie $h$-invariant level-families of this ansatz live on the $Lie$
$h$-$invariant$ $graphs$, whose Lie group-theoretic characterization is
obtained in Section 7.4.
The Lie characterization of these graphs is another facet of the underlying Lie
group structure of graph theory.

In parallel with the conformal field-theoretic development, we also
characterize:

a) The $Lie$ $h$-$invariant$ $graph$ $multiplets$.

b)  K$_{g/h}$-$complementarity$ through the $g/h$ coset graphs, and other
    generalized graph complementarities which correspond to the generalized
    K-conjugations.

c)  The $self$ K$_{g/h}$-$complementary$ $graphs$.

\noindent
Examples and some enumeration of these new graph categories are given in
Sections 7.5 and 10, and some exact solutions on Lie $h$-invariant graphs are
given in Section 11.

It would be interesting to apply the general theory of Lie $h$-invariant
conformal constructions to the graph theory of superconformal
level-families \cite{29,30}, and to the other generalized graph theories of the
Virasoro and superconformal master equations.

\bigskip\noindent
{\large\bf 2.} {\large\bf The Virasoro Master Equation}
\bigskip

The general affine-Virasoro construction begins with the currents
$J_{a}$ of untwisted affine Lie $g$ \cite{1,2}
$$ [J_{a}^{(m)},J_{b}^{(n)}]=i{f_{ab}}^{c}J_{c}^{(m+n)}+mG_{ab}\d_{m+n,
0} \eqno(2.1a)$$
$$a,b=1,\cdots,{\fam=0\tenrm dim}\,g\,\,\,\,,\,\,\,\,m,n\,\in\,\Z
\eqno(2.1b)$$
where ${f_{ab}}^{c}$ and $G_{ab}$ are respectively the structure constants and
general Killing metric of $g=\oplus_{I}g_{I}$.
To obtain level $x_{I}=2k_{I}/\psi_{I}^{2}\,\,$ of $g_{I}$
with dual Coxeter number ${\tilde h}_{I}=Q_{I}/\psi_{I}^{2}$, take
$$ G_{ab}=\oplus_{I}k_{I}\eta^{I}_{ab}\,\,\,\,,\,\,\,\,
{f_{ac}}^{d}{f_{bd}}^{c}=-\oplus_{I}Q_{I}\eta^{I}_{ab}\eqno(2.2)$$
where $\eta^{I}_{ab}$ and $\psi_{I}$ are respectively the Killing metric and
highest root of $g_{I}$.
The class of operators
$$ T(L)\equiv L^{ab}\xx J_{a}J_{b}\xx \equiv \sum_{m\in \Z} L^{(m)
}z^{-m-2}\eqno(2.3)$$
is defined with symmetric normal ordering, $T_{ab}\equiv \xx J_{a}J_{b}\xx =
T_{ba}$, and $L^{ab}=L^{ba}$ is called the inverse
inertia tensor, in analogy with the spinning top.
In order that $T(L)$ be a Virasoro operator
$$ [L^{(m)},L^{(n)}]=(m-n)L^{(m+n)}+{c\over 12}m(m^{2}-1)\d_{m+n,0}\eqno(2.4)$$
the inverse inertia tensor must satisfy the Virasoro master equation (VME)
 \cite{15,16}
$$L^{ab}=R^{ab}(L)\eqno(2.5a)$$
$$R^{ab}(L)=2L^{ac}G_{cd}L^{db}-L^{cd}L^{ef}{f_{ce}}^{a}{f_{df}}^{b}-L^{cd}
{f_{ce}}^{f}{f_{df}}^{(a}L^{b)e}\eqno(2.5b)$$
$$c(L)=2G_{ab}L^{ab}\eqno(2.5c)$$
where we have defined $A_{(a}B_{b)}\equiv A_a B_b + A_b B_a$.
The VME has been identified \cite{34} as an Einstein-like system
on the group manifold with central charge $c=$dim$g-4R$, where $R$ is the
Einstein curvature scalar.

We remark on some general properties of the master equation which will be
useful below.

1. The affine-Sugawara construction \cite{2,5,9} $L_{g}$ is
$$ L_{g}^{ab}=\oplus_{I}{\eta_{I}^{ab}\over 2k_{I}+Q_{I}}\,\,\,,\,\,\,
c_{g}=\sum_{I}{x_{I}{\fam=0\tenrm dim}\,g_{I}\over x_{I}+{\tilde h}_{I}}
\eqno(2.6)$$
for arbitrary levels of affine $g$, and similarly for $L_{h}$ when $h\subset
g$.

2. K-conjugation covariance \cite{2,5,10,15}.
When $L$ is a solution of the master equation on $g$, then so is the
K-conjugate partner $\tilde{L}$ of $L$
$$\tilde{L}^{ab}=L^{ab}_{g}-L^{ab}\,\,\,,\,\,\,c(L)=c_{g}-c(\tilde{L}) \eqno(2.
7)$$
while the corresponding stress tensors $T(L)$ and $T(\tilde{L})$ form a
commuting pair of Virasoro operators.
The coset constructions \cite{2,5,10} $L_{g/h}=L_{g}-L_{h}$ are the K-conjugate
partners of $L_{h}$ on $g$.

3. Affine-Sugawara nests \cite{20}. Repeated K-conjugation on the embedded
   subgroup sequence $g\supset h_{1}\supset\cdots\supset h_{d}$ generates an
   affine-Sugawara nest of depth $d+1$
$$L_{g/h_{1}/h_{2}\cdots /h_{d}}=L_{g}-L_{h_{1}}+L_{h_{2}}-\cdots +(-1)
^{d}L_{h_{d}}\eqno(2.8a)$$
$$c_{g/h_{1}/h_{2}\cdots /h_{d}}=c_{g}-c_{h_{1}}+c_{h_{2}}-\cdots +(-1)
^{d}c_{h_{d}}\eqno(2.8b)$$
so that the affine-Sugawara and coset constructions are affine-Sugawara nests
of depth 1 and 2 respectively.
The more general affine-Virasoro nests \cite{20} are formed from (2.8) by the
replacement $L_{h_{d}}\rightarrow L(h_{d})$, where $L(h_{d})$ is any
construction on $h_{d}$.

4. Unitarity \cite{10,20}. Unitary constructions on positive integer level of
affine compact $g$ are recognized when $L^{ab}=$real  in any Cartesian basis,
and corresponding forms in other bases.

5. Automorphisms \cite{34,24,25}. The elements $\o\in Aut\,g$ of the
   automorphism group of $g$ satisfy
$$  {f_{ab}}^c= {\o_a}^d {\o_b}^e {(\o^{-1})_f}^c {f_{de}}^f \eqno(2.9a)$$
$$  G_{ab}={\o_a}^c {\o_b}^d G_{cd} \eqno(2.9b)$$
and (2.9b) may be written as
$${(\o^{T})_{a}}^{b}\equiv G^{bc}{\o_{c}}^{d}G_{da}={(\o^{-1})_{a}}^{b}
\eqno(2.10)$$
so that $\o$ is a (pseudo) orthogonal matrix which is an element of
$SO(p,q)$, $p+q=$dim$\,g$ with metric $G_{ab}$. The automorphism
group includes the outer automorphisms of $g$ and the inner automorphisms
$$  g(\o) T_a g^{-1}(\o) ={\o_a}^b T_b \;\,,\;\;\;g(\o)\, \in\, G
\eqno(2.11)$$
where $T$ is any representation of $G$.

The transformation
$$  {J'_a}^{(m)} ={\o_a}^b J_b^{(m)}\,\,\,,\,\,\,\o\,\in\,Aut\,g\eqno(2.12)$$
is an automorphism of affine Lie $g$ in (2.1), and $(L')^{ab}$ defined by
$$  (L')^{ab}=L^{cd} {(\o^{-1})_c}^a  {(\o^{-1})_d}^b \eqno(2.13a)$$
$$  R(L')^{ab}=R^{cd}(L){(\o^{-1})_c}^a  {(\o^{-1})_d}^b\,\,\,,\,\,\,\o\,\in\,
Aut\,g\eqno(2.13b)$$
is an automorphically equivalent solution of the VME when $L^{ab}$ is a
solution.
The corresponding matrix form of the automorphism (2.13)
$$L'=\o L\o^{-1}\,\,\,,\,\,\,{L_{a}}^{b}\equiv G_{ac}L^{cb}\eqno(2.14a)$$
$$R(L')=\o R(L)\o^{-1}\,\,\,,\,\,\,{R_{a}}^{b}\equiv G_{ac}R^{cb}\,\,\,,\,\,\,
\o\,\in\,Aut\,g\eqno(2.14b)$$
is preferred below.

6. Self K-conjugate constructions \cite{25,26,33}. These constructions satisfy
$${\tilde L}=\o L\o^{-1}\,\,\,\,\,\,\hbox{for some}\;\o\,\in\,Aut\,g\eqno(
2.15a)$$
$$c={c_{g}\over 2}\eqno(2.15b)$$
on Lie group manifolds of even dimension.
The first examples were observed on $SO(4n)$ and $SO(4n+1)$, where they
correspond to the self-complementary graphs of graph theory, and they have also
been observed on $SU(3)$ and $SU(5)$.

\bigskip\noindent
{\large\bf 3.} {\large\bf Group-Invariant Ans\"atze}
\bigskip

As in general relativity, consistent
ans\"atze\footnote{A consistent ansatz \cite{20} of the VME preserves an
equal number $n$ of quadratic equations and unknowns, so it
contains $2^{n}$ level-families generically.}
and subans\"atze
have played a central role in obtaining exact solutions of the VME
on affine $g$.
In this section, we study the broad class of consistent ans\"atze associated
to symmetry under subgroups of $Aut\,g$.

In the VME on affine $g$, the $H$-invariant ansatz $A_{g}(H)$ is
$$A_{g}(H)\,\,\,:\,\,\,\,L=\o L\o^{-1}\,\,\,,\,\,\,\forall\,\,\o\,\in\,H\,
\subset\, Aut\,g\eqno(3.1)$$
where $H$ may be any finite subgroup or Lie subgroup of $Aut\,g$, and may
involve inner or outer automorphisms of $g$.
The ansatz $A_{g}(H)$ is a set of linear relations on $L^{ab}$ which requires
that all the conformal field theories (CFTs) of the ansatz are invariant under
$H$.
The ansatz is consistent because , according to (2.13), the same linear
relations
$$R(L)=R(\o L\o^{-1})=\o R(L)\o^{-1}\,\,\,,\,\,\forall\,\o\,\in\,H\,\subset\,
Aut\,g\eqno(3.2)$$
are obtained for the right-hand side of the VME.

As a set, the $H$-invariant ans\"atze on $g$ follow the subgroup embeddings in
$Aut\,g$, so that
$$  A_{g}(\au) \subset A_{g}(H_1) \subset A_{g}(H_2) \cdots \subset A_{g}(H_n)
 \subset A_{g}({\tenbf 1})\eqno(3.3a)$$
$$\au \supset H_1 \supset H_2 \cdots \supset H_n \supset {\bf 1}\eqno(3.3b)$$
where $\bf 1$ is the trivial subgroup.
Subgroup embedding in $\au$ is a formidable problem, but we may
discuss many examples.

The largest $H$-invariant ansatz $A_{g}({\bf 1})$ is the VME itself on affine
$g$.
The smallest $H$-invariant ansatz $A_{g}(\au)$ contains no new constructions
because it resides in the smallest inner automorphic ansatz
$$A_{g}(G)\,\,\,:\;\,\,\,\,L=\o L\o^{-1}\,\,\,,\,\,\,\forall\,\o\,\in\,G
\eqno(3.4)$$
which itself contains no new constructions: The explicit form of $A_{g}(G)$ on
$G=\otimes_{I}G_{I}$ is
$$A_{g}(G)\,\,\,:\,\,\,\,L^{ab}=\oplus_{I}L_{I}\eta_{I}^{ab}\eqno(3.5)$$
and the solutions of $A_{g}(G)$, which are all possible $G$-invariant CFTs on
affine $g$, are $L=0$ and the affine-Sugawara constructions on all subsets of
$g_{I}$.

Here are some less trivial examples.

1. The graph-symmetry subans\"atze in the graph-theory ansatz
$SO(n)_{diag}$ \cite{25}.
These H-invariant subans\"atze are reviewed in Section 7.2.

2. Outer automorphism. The Chevalley involution on simple $g$,
$$  H'_A=-H_A\;\,,\;\;\;\;\;E_{\a}'=-E_{-\a}\eqno(3.6a)$$
$$  {\o_a}^b=-\m_{ab}=-\left( \matrix{ \d_{AB} & 0 \cr 0 & \d_{\a+\b,0} }
\right) \;\,,\;\;\;\;\forall\,A\, \in\, CSA\,\,,\,\,\a\, \in \,\Phi(g)\eqno(3.
6b)$$
is an involutive automorphism of $g$ which is an outer automorphism for
$SU(n \geq 3)$, $SO(2n \geq 6 )$ and $E_6$.
The corresponding consistent ``complex conjugation"-invariant ansatz
$$L^{A\a}=L^{A,-\a}\,\,\,,\,\,\,L^{\a\b}=L^{-\a,-\b}\,\,\,,\,\,\,{\rm no}\,\,
{\rm restriction}\,\,{\rm on}\,\,L^{AB}\eqno(3.7)$$
follows from (3.1) and (3.6).

3. Inner automorphism. The ``grade automorphism" of simple $g$
$$H'_{A}=H_{A}\,\,\,,\,\,\,E'_{\a}=e^{{i\pi\over h}G(\a)}E_{\a}\eqno(3.8a)$$
$$  {\o_a}^b =\left( \matrix{ \d_{AB} & 0 \cr 0 & e^{{i\pi\over h}G(\a)}
\d_{a,\b} }\right)\eqno(3.8b)$$
generates a cyclic subgroup of $\au$, where $h$ is the Coxeter number of $g$
and
$G(\a)$ is the grade of $\a$ $\in$ $\Phi(g)$ (grade=number of simple roots in
$\a$).
The corresponding consistent invariant ansatz has non-zero components
$$L^{AB}\,\,\,,\,\,\,L^{\a\b}\,\,\,{\rm when}\,\,\,G(\a)+G(\b)=0\,.\eqno(3.9)$$

4. A metric ansatz. Consider $SU(n)$ in the Pauli-like basis \cite{35}
$$ \m_{\bp,\bq}=\s(\bp)\d_{\bp,(-\bq)({\rm mod}\,n)} \eqno(3.10a)$$
$$ {f_{\bp,\bq}}^{\br}=-\sqrt{{2 \psi_{n}^{2} \over n}}\s(\bp,\bq)
\sin(\frac{\pi}{n}(\bp \times \bq))\d_{(\bp+\bq)({\rm mod}\,n),\br}
\eqno(3.10b)$$
where the adjoint index is $a={\vec p}=(p_{1},p_{2})$ with $p_{1}$,$p_{2}$
integers between $0$ and $n-1$, excluding ${\vec p}={\vec 0}$.
The phases $\sigma({\vec p})$ and $\sigma({\vec p},{\vec q})$ are given in Ref.
[35].
It is not difficult to check that
$$  {\o_{\bp}}^{\bq \,\,(1)} = e^{i2\pi p_1/n}\d_{\bp,\bq}\,\,\,\,,\,\,\,\,
{\o_{\bp}}^{\bq \,\,(2)} = e^{i2\pi p_2/n}\d_{\bp,\bq} \eqno(3.11)$$
are automorphisms which generate the finite subgroup $Z_{n}\times Z_{n}$ of
$Aut\,SU(n)$.
The corresponding consistent invariant ansatz
$$L^{{\vec p},{\vec q}}=L_{\vec p}\eta^{{\vec p},{\vec q}}\eqno(3.12)$$
is the metric ansatz $SU(n)_{\rm metric}$$^{26,28}$ on $SU(n)$,
whose unitary irrational level-families are classified by the sine-area
graphs \cite{26,27,28}.

\bigskip\noindent
{\large\bf 4.} {\large\bf Lie {\bfi h}-Invariant Ans\"atze}
\bigskip

When $H\subset G$ is a connected Lie subgroup, the ansatz $A_{g}(H)$ in (3.1)
is equivalently described by its infinitesimal form
$$A_{g}({\rm Lie}\,h)\,\,\,:\,\,\,\, \d L^{ab}(\psi)=L^{c(a}{f_{cd}}^{b)}
\psi^d =0\eqno(4.1)$$
where $\psi$ parametrizes $H$ in the neighborhood of the origin
$$  {(\o)_a}^b=\d_a^b+\psi^c{f_{ca}}^b +{\cal O}(\psi^2) \,\,\,,\,\,\,\forall\,
\o\,\in\,H\eqno(4.2)$$
and $\delta L=\o L\o^{-1}-L+{\cal O}(\psi^{2})$ is an infinitesimal
transformation of $L$ in $H$.
The ans\"atze $\{A_{g}({\rm Lie}\,h)\}$ will be called the $Lie$
$h$-$invariant$
$ans\ddot atze$ because all the CFTs of $A_{g}({\rm Lie}\,h)$ have at least
the Lie $h$ symmetry (4.1).
The generic Lie $h$-invariant ansatz is, like the VME itself, a large set of
coupled quadratic equations, so we expect that the generic Lie $h$-invariant
construction has irrational central charge.

Since the affine-Sugawara construction $L_{g}$ in (2.6) is Lie $h$-invariant
for all $h\subset g$, it follows from (4.1) that the K-conjugate partner
${\tilde L}=L_{g}-L$ of a Lie $h$-invariant construction $L$ is also Lie
$h$-invariant.
Put another way, every Lie $h$-invariant ansatz is K-conjugation covariant.

The Lie $h$-invariant ans\"atze follow the Lie subalgebra
embeddings of $g$, as in Eq. (3.3). Some examples of $A_{g}($Lie $h)$ are:

1. $A_{g}(CSA)$. The non-vanishing components of the Cartan
   subalgebra-invariant ansatz
$$A_{g}(CSA)\,\,\,:\,\,\,\,L^{AB}\,\,\,,\,\,\,L^{\a,-\a}\,\,\,,\,\,\,\,A\,\in\,
CSA\,\,\,,\,\,\,\a\,\in\,\Phi(g)\eqno(4.3)$$
are obtained from (4.1) by setting $\psi^{\alpha}=0$ and $\psi^{A}$ arbitrary.

2. $A_{SU(3)}($Lie $h)$. Up to automorphisms, the complete list of Lie
   $h$-invariant ans\"atze on $SU(3)$ is
$$  A_{SU(3)}(U(1)): \left\{ \matrix{
L^{11}=L^{22},\;\,L^{44}=L^{55}, \;\,L^{66}=L^{77} \cr
L^{46}=-L^{57},\;\,L^{47}=L^{56}\cr
 L^{33},\;\,L^{88},\;\,L^{38}}\right.\eqno(4.4)$$
$$  A_{SU(3)}(U(1)\times U(1)): \left\{ \matrix{
L^{11}=L^{22},\;\,L^{44}=L^{55},\;\,L^{66}=L^{77} \cr
 L^{33},\;\,L^{88},\;\,L^{38} } \right.\eqno(4.5)$$
$$ A_{SU(3)}(SU(2)_{reg})=A_{SU(3)}(SU(2)_{reg}\times U(1)):  \left\{
\matrix{ L^{11}=L^{22}=L^{33},\;\,L^{88} \cr L^{44}=L^{55}=L^{66}=L^{77} }
\right.\eqno(4.6)$$
$$ A_{SU(3)}(SU(2)_{irreg}):\left\{ \matrix{
 L^{11}=L^{33}=L^{44}=L^{66}=L^{88} \cr L^{22}=L^{55}=L^{77}\,\,\,. }\right.
\eqno(4.7) $$
This set of ans\"atze was obtained by choosing
$$U(1)\sim J_{3}\,\subset\,U(1)\times U(1)\sim J_{3},J_{8}\subset SU(2)
_{reg}\times U(1)\sim J_{1},J_{2},J_{3},J_{8}\eqno(4.8a)$$
$$SU(2)_{irreg}\sim J_{2},J_{5},J_{7}\eqno(4.8b)$$
in the Gell-Mann basis, and automorphic copies of these ans\"atze are obtained
by choosing other representatives of the subgroups

Here is what we know about these ans\"atze:

a) $A_{SU(3)}(SU(2)_{irreg})$. All 4 level-families of this ansatz are
affine-Sugawara nests: $L=0$, $SU(2)_{4x}$ and their K-conjugates on $SU(3)
_{x}$.

b) $A_{SU(3)}(SU(2)_{reg})$. All 8 level-families of this ansatz are
affine-Sugawara nests: $L=0$, $SU(2)_{x}$, $U(1)$, $SU(2)_{x}\times U(1)$
and their K-conjugates on $SU(3)_{x}$.

c) $A_{SU(3)}(U(1)\times U(1))$. This is a subansatz of the ansatz $SU(3)
_{BASIC}$, which has been completely solved \cite{24}.
We have checked that $SU(3)_{BASIC}$, and hence $A_{SU(3)}$
$(U(1)\times U(1))$,
contains no ($U(1)\times U(1)$)-invariant level-families beyond the relevant
affine-Sugawara nests, Cartan $SU(3)^{\#}$ and $SU(3)/$Cartan $SU(3)^{\#}$
\cite{20}.

d) $A_{SU(3)}(U(1))$. We have not obtained all 256 level-families of this
ansatz,
but it contains at least one automorphic copy of each of the 8 known $U(
1)$-invariant\footnote{The representatives of (4.9a) and (4.9b) in Ref. [24]
have $U(1)\sim J_{5}$
and $U(1)\sim(\sqrt{3}J_{3}-J_{8}/2$ respectively.}
level-families with unitary irrational central charge \cite{23,24}
$$SU(3)_{BASIC}\supset\matrix{ SU(3)^{\#}_{D(1)},
SU(3)/SU(3)^{\#}_{D(1)}, SU(3)^{\#}_{D(2)}, SU(3)/SU(3)^{\#}_{D(2)}\;(4.9a)
\cr
SU(3)^{\#}_{A(1)}, SU(3)/SU(3)^{\#}_{A(1)}, SU(3)_{A(2)}^{\#},
SU(3)/SU(3)_{A(2)}^{\#}.\;(4.9b)}$$

3. $A_{SO(n)}(SO(m))$. According to (1.5), the generic number of conformal
   level-families in the VME on affine $SO(n)$ is
$$N(SO(n))=2^{n(n-2)(n^{2}-1)/8}={\cal O}(2^{n^{4}/8})\,\,.\eqno(4.10)$$
Among these, we estimate the number of $SO(m)$-invariant level-families as
follows.
The standard Cartesian basis of $SO(n)$ has adjoint index
$$a=ij\,\,\,\,,\,\,\,\,1\leq i < j\leq n\eqno(4.11)$$
where $i$ and $j$ are vector indices of $SO(n)$, and the inverse inertia tensor
is labeled $L^{ij,kl}$ in this basis. Let
$$i=(\mu,I)\,\,\,,\,\,\,\mu=1,2,\cdots,m\,\,\,,\,\,\,I=m+1\cdots,n\eqno(4.12)$$
so that $\mu$ is a vector index of $SO(m)$.
The $SO(m)$-invariant components of the inverse inertia tensor
$$\sum_{\mu\nu}L^{\mu\nu,\mu\nu}\,\,\,,\,\,\,\sum_{\mu}L^{\mu I,\mu I}\,\,\,,\,
\,\,L^{IJ,KL}\eqno(4.13)$$
may be taken as the non-zero unknowns of the ansatz $A_{SO(n)}(SO(m))$, so we
count
$$|A_{SO(n)}(SO(m))|=2^{1+{(n-m)(n-m+1)((n-m)(n-m-1)+6)\over 8}}={\cal O}(
2^{n^{4}/8})\,\,{\rm at}\,\,{\rm fixed}\,\,m\eqno(4.14)$$
$SO(m)$-invariant level-families in the VME on $SO(n)$, where the exponent in
(4.14) is the number of unknowns in (4.13).
The number (4.14) has not been corrected for residual continuous or
discrete automorphisms of the ansatz, but this estimate and the corresponding
Lie $h$-invariant fraction
$${|A_{SO(n)}(SO(m))|\over N(SO(n))}={\cal O}(2^{-n^{3}m/2})\,\,{\rm at}\,\,
{\rm fixed}\,\,m\eqno(4.15)$$
provide a strong indication that Lie $h$-invariant
constructions, although not generic, are copious in the space of CFTs.
A more precise statement of these conclusions, including a comparison with
the number of affine-Sugawara nests, is obtained with graph theory in
Section 7.5.

\bigskip\noindent
{\large\bf 5.} {\large\bf (0,0) and (1,0) Operators in Lie {\bfi h}-Invariant
CFTs}

The simplest examples of Lie $h$-invariant CFTs on $g$ are the affine-Sugawara
nests, which include the affine-Sugawara constructions, the coset construction
and the coset nests.
Each of these contains certain numbers $N_{0}$ and $N_{1}$ of (0,0) and (1,0)
operators among the currents of affine $g$.
As examples, consider
$$\matrix{L_{g}\subset A_{g}({\rm Lie}\,h)\hfill&:&N_{0}=0\hfill&,&
N_{1}={\rm dim}\,g\hfill\cr
L_{h}\subset A_{g}({\rm Lie}\,h)\hfill&:&N_{0}={\rm dim}\,h'\hfill&,&
N_{1}={\rm dim}\,h\hfill\cr
L_{g/h}\subset A_{g}({\rm Lie}\,h)\hfill&:&N_{0}={\rm dim}\,h\hfill&,&
N_{1}={\rm dim}\,h'\hfill\cr}\eqno(5.1)$$
where $h'\subset g$ is the centralizer of $h$ in $g/h$ (see Ref. [12]).
In the case of non-trivial $h'$, the subgroup and coset constructions have the
higher Lie symmetry $h\oplus h'$, and also appear in $A_{g}(h\oplus h')$.

The general affine-Sugawara nest of depth $d+1$
$$L_{g/h_{1}/\cdots/h_{d}}\,\,\subset \,\,A_{g}({\rm Lie}\,h_{d})\eqno(5.2)$$
has at least a Lie $h_{d}$-symmetry, which is generated by the current
subalgebra $\{J_{a}$, $a\,\in\,h_{d}\}$.
This set of currents is (0,0) or (1,0) of the nest when $d$ is odd or even
respectively.

Consistent with this data, Appendix A contains the proof of the following
general results.

\u{THEOREM}. Let $L($Lie $h)\in A_{g}($Lie $h)$ be a unitary
conformal construction whose Lie symmetry is exactly $h$.
Then
$$N_{0}+N_{1}={\rm dim}\,h\eqno(5.3)$$
where $N_{0}$, $N_{1}$ is the number of (0,0) and (1,0) currents in the
construction, and the Lie $h$ symmetry of $L($Lie $h)$ is generated by this set
of currents.

\u{REMARK}. This theorem was stated without proof in Ref. [33].

\u{PROPOSITION}. The symmetry algebra Lie $h$ of $L($Lie $h)$
has the form
$$h=h_{0}\oplus h_{1}\,\,\,,\,\,\,{\rm dim}\,h_{0}=N_{0}\,\,\,,\,\,\,{\rm
dim}\,h_{1}=N_{1}\,.\eqno(5.4)$$

\u{REMARK}. Since
$$[L^{(m)}({\rm Lie}\,h),J_{A}^{(n)}]=0\,\,\,,\,\,\,\forall\,n\,\,,\,\,A\,\in\,
h_{0}\eqno(5.5a)$$
$$[L^{(m)}({\rm Lie}\,h),J_{I}^{(0)}]=0\,\,\,,\,\,\,I\,\in\,h_{1}\eqno(5.5b)$$
we will call $h_{0}$ and $h_{1}$ the affine (or local) and global components of
the Lie $h$ symmetry respectively.
It also follows from (5.5) that K-conjugation interchanges the global and local
components of Lie $h$
$$h_{0}({\tilde L})=h_{1}(L)\,\,\,\,,\,\,\,\,h_{1}({\tilde L})=h_{0}(L)\eqno(5.
6)$$
where ${\tilde L}($Lie $h)$ is the K-conjugate partner  of $L($Lie $h)$.

\bigskip\noindent
{\large\bf 6.} {\large\bf The Lie {\bfi h}-Invariant Conformal Multiplets}
\bigskip

In what follows, we refer to ordinary K-conjugation on affine $g$ as
K$_{g}$-conjugation, in order to distinguish this operation from the
generalized
K-conjugations discussed below.
We also visualize the K$_{g}$-conjugate partners $L$ and ${\tilde L}$ as a
K$_{g}$-doublet
$$ \matrix{_{\phantom{g}}L_{\phantom{g}} \cr \mapupdown{{\rm K}_g} \cr
{\tilde L}=L_g-L }\eqno(6.1)$$
which is closed under K$_{g}$-conjugation because K$_{g}^{2}=1$.

When a conformal construction $L({\hat h}$ on $g)$ on affine $g$ has an affine
(or $\hh$) invariance
$$[L^{(m)}({\hat h}\,\,{\rm on}\,\,g)\,,\,J_{A}^{(n)}]=0\,\,\,,\,\,\,A\,\in\,
h\eqno(6.2)$$
then it resides in the $Lie$ $h$-$invariant$ $quartet$

$$  \matrix{ L(\hat{h}\; {\rm on}\; g) & \mapright{+L_h} &
L(\hat{h}\; {\rm on}\; g) + L_h \cr
\mapupdown{{\rm K}_g} &  & \mapupdown{{\rm K}_g} \cr
L_g-L(\hat{h}\; {\rm on}\; g) & \mapleft{+L_h}
& L_{g/h}-L(\hat{h}\; {\rm on}\; g) } \eqno(6.3a)$$
\bigskip

$$  \matrix{ c & \mapright{+L_h} & c+c_h \cr
\mapupdown{{\rm K}_g} &  & \mapupdown{{\rm K}_g} \cr
c_g-c & \mapleft{+L_h} & c_{g/h}-c  }\eqno(6.3b)$$

\noindent
where $L_{h}$ is the affine-Sugawara construction on $h$ and
$L_{g/h}=L_{g}-L_{h}$ is the $g/h$ coset construction.
All 4 members of the quartet are Lie $h$-invariant conformal constructions
because $T(L_{h})$ commutes with $T(L({\hat h}$ on $g))$, and all 4 members
of the quartet occur in the ansatz $A_{g}($Lie $h)$.
Note, however, that only two of the constructions in the quartet
$$L({\hat h}\;{\rm  on}\; g)\,\,\,\,,\,\,\,\,L_{g/h}-L({\hat h}\;{\rm on}\;g)
\eqno(6.4)$$
are $\hh$-invariant, while the other two constructions
$$L({\hat h}\;{\rm on}\;g)+L_{h}\,\,\,\,,\,\,\,\,L_{g}-L({\hat h}\;{\rm on}\;g)
\eqno(6.5)$$
are globally $h$-invariant.

Looking more closely, we see that two generalized K-conjugations are
defined in the quartet.
In the first place, there is a K$_{g/h}$-$conjugation$ through
the $g/h$ coset construction
$$  \matrix{ L(\hat{h}\; {\rm on}\; g) &  &
L(\hat{h}\; {\rm on}\; g) + L_h \cr
 & \mapdiagr{{\rm K}_{g/h}} &  \cr L_g-L(\hat{h}\; {\rm on}\; g) & & L_{g/h}
-L(\hat{h}\; {\rm on}\; g) }\eqno(6.6)$$

\noindent
which acts with K$_{g/h}^{2}=1$ in the space of $\hh$-invariant
CFTs.
Algebraically, we may write that the K$_{g/h}$-conjugate partner $({\tilde L})
_{g/h}$ of an $\hh$-invariant conformal construction $L$
$${\rm K}_{g/h}\;\;:\;\;({\tilde L})_{g/h}\equiv L_{g/h}-L\,\,\,,\;\,\,\,
({\tilde c})_{g/h}=c_{g/h}-c\eqno(6.7)$$
is also an $\hh$-invariant conformal construction.
We also see a K$_{g+h}$-$conjugation$ ($\equiv $ first K$_{g}$
and then $+L_{h}$)
$$  \matrix{ L(\hat{h}\; {\rm on}\; g) & &
L(\hat{h}\; {\rm on}\; g) + L_h \cr
&\mapdiagl{{\rm K}_{g+h}}  & \cr
L_g-L(\hat{h}\; {\rm on}\; g) &
& L_{g/h}-L(\hat{h}\; {\rm on}\; g) } \eqno(6.8)$$
which acts with K$_{g+h}^{2}=1$ in the space of globally $h$-invariant
CFTs.
In this case, the K$_{g+h}$-conjugate partner $({\tilde L})_{g+h}$
of a globally $h$-invariant conformal construction $L$
$${\rm K}_{g+h}\;\;:\;\;({\tilde L})_{g+h}\equiv (L_{g}+L_{h})-L\;\;,\;\;\;
({\tilde c})_{g+h}=(c_{g}+c_{h})-c\eqno(6.9)$$
is also a globally $h$-invariant conformal construction,
although $L_{g}+L_{h}$ is not a conformal construction.

The simplest example of a Lie $h$-invariant quartet is the familiar set
$$ \matrix{ L=0 & \mapright{+L_h} & L_h \cr
\mapupdown{{\rm K}_g} &  & \mapupdown{{\rm K}_g} \cr
L_g & \mapleft{+L_h} & L_{g/h}} \eqno(6.10)$$
obtained by choosing $L({\hat h}$ on $g)=0$.
Moreover, the 8 known unitary $U(1)$-invariant level-families with irrational
central charge \cite{23,24}
$$\eqalignno{&\matrix{ SU(3)_{D(1)}^{\#} & \mapright{+L_{U(1)}} &
SU(3)_{D(2)}^{\#}\cr
\mapupdown{{\rm K}_{SU(3)}} &  & \mapupdown{{\rm K}_{SU(3)}} \cr
SU(3)/SU(3)_{D(1)}^{\#}& \mapleft{+L_{U(1)}}
& L_{SU(3)/U(1)}-SU(3)_{D(1)}^{\#}\cr}&(6.11a)\cr
&\cr &\cr
&\matrix{ SU(3)_{A(1)}^{\#} & \mapright{+L_{U(1)}} &
SU(3)_{A(2)}^{\#}\cr
\mapupdown{{\rm K}_{SU(3)}} &  & \mapupdown{{\rm K}_{SU(3)}} \cr
SU(3)/SU(3)_{A(1)}^{\#}& \mapleft{+L_{U(1)}}
& L_{SU(3)/U(1)}-SU(3)_{A(1)}^{\#}\cr}&(6.11b)\cr}$$
form 2 $U(1)$-invariant quartets, as shown.

In analogy with the self K$_{g}$-conjugate constructions \cite{25,26,33} in
(2.15), it is natural to expect the occurrence of $self$ K$_{g/h}$-$conjugate$
$constructions$
$$({\tilde L})_{g/h}=L_{g/h}-L({\hat h}\;{\rm on}\;g)=\o L({\hat h}\;{\rm on}\;
g)\o^{-1}\,\,\,,\,\,\,\o\,\in\,Aut\,g\eqno(6.12a)$$
$$({\tilde c})_{g/h}=c={c_{g/h}\over 2}\eqno(6.12b)$$
in which K$_{g/h}$-conjugate partners are automorphically equivalent.
Note that self K$_{g/h}$-conjugate constructions will occur with $half$
$coset$ $central$ $charge$ because $c_{g/h}-c=c$.
The existence of these constructions is established with
graph theory
in Section 10.2, and exact self K$_{g/h}$-conjugate level-families
are obtained on $SO(6)$ in Section 11.4.

Moreover, self K$_{g/h}$-conjugate constructions occur in conjunction
with $self$ K$_{g+h}$-$conjugate$ $constructions$
$${\tilde L}({\hat h}\;{\rm on}\;g)=L_{g}-L({\hat h}\;{\rm on}\;g)=\o^{-1}
(L({\hat h}\;{\rm on}\;g)+L_{h})\o\,\,\,,\,\,\,\o\,\in\,Aut\,g\eqno(6.13a)$$
$${\tilde c}=c+c_{h}={c_{g}+c_{h}\over 2}\eqno(6.13b)$$
because (6.13a) follows from (6.12a).

The Lie $h$-invariant quartets are only the first glimpse into a web of $Lie$
$h$-$invariant$ $conformal$ $multiplets$, associated with multiple subgroup
embeddings.
As an example, the embedding $g\supset h\supset h'$ underlies the Lie
$h'$-invariant\footnote{The $h$-invariant quartet of $L(\hh$ on $g)$ is the top
face of the cube, while the lower members of the multiplet are only
$h'$-invariant generically. The entire octet
is $h$-invariant in the special case where
$h=h_{1}\oplus h_{2}$ and $h'=h_{1}$ or $h_{2}$.}
conformal octet shown in Fig. 1, which is constructed from
$L({\hat h}\;{\rm on}\;g)$ by the moves K$_{g}$, $+L_{h}$ and $+L_{h'}$.
A number of new K-conjugations are observed in the octet, including
K$_{g/h/h'}$-$conjugation$ through the affine-Sugawara nest of depth 3,
$${\rm K}_{g/h/h'}\,\,\,:\,\,\,(L')_{g/h/h'}\equiv L_{g/h/h'}-
L({\hat h}\;{\rm on}\;g)\,\,.\eqno(6.14)$$
\noindent
Examples of these octets are given in Section 10.1.

More generally, a $2^{d+1}$-plet of Lie $h_{d}$-invariant CFTs is associated to
the general subgroup nest $g\supset h_{1}\cdots \supset h_{d}$ of depth $d+1$.
These $(d+1)$-dimensional cubes may be generated  from $L({\hat h}_{1}\;
{\rm on}\;g)
$ by the moves K$_{g}$ and $+L_{h_{i}}$, $i=1,\cdots,d$, and the multiplets
show a broad variety of generalized K-conjugations, including
K$_{g/h_{1}/\cdots/h_{d}}$-$conjugation$ through the general affine-Sugawara
nest.
We conjecture that generalized self K-conjugate constructions exist for all
the generalized
K-conjugations, including $self$ K$_{g/h_{1}/\cdots/h_{d}}$-$conjugate$
$constructions$ with central charge
$$c={1\over 2}c_{g/h_{1}/\cdots/h_{d}}\eqno(6.15)$$
but we will not investigate these higher multiplets in this paper.

We finally remark that the Lie $h$-invariant multiplets are special cases of
a more general class of conformal multiplets associated to affine-Virasoro
nesting \cite{20}.
Consider, for example, the conformal quartet
$$  \matrix{ L(\hat{h}\; {\rm on}\; g) & \mapright{+L(h^{\#})} &
L(\hat{h}\; {\rm on}\; g) + L(h^{\#}) \cr
\mapupdown{{\rm K}_g} &  & \mapupdown{{\rm K}_g} \cr
L_g-L(\hat{h}\; {\rm on}\; g) & \mapleft{+L(h^{\#})}
& L_{g/h^{\#}}-L(\hat{h}\; {\rm on}\; g) } \eqno(6.16)$$
where $L(h^{\#})$ is $any$ construction on affine $h$.
These quartets show K$_{g/h^{\#}}$-$conjugation$ through the simple
affine-Virasoro nest $L_{g/h^{\#}}=L_{g}-L(h^{\#})$,
but the constructions on the right of (6.16) carry only the Lie symmetry of $L(
h^{\#})$, which is generically asymmetric.

\bigskip\noindent
{\large\bf 7.} {\large\bf Lie Symmetry in Graph Theory}

\bigskip
\noindent
{\bf 7.1} {\bf Strategy}
\bigskip

The generically unitary and irrational conformal level-families of the ansatz
$SO(n)_{diag}$ \cite{25} are isomorphic to the
graphs of order $n$, and the Lie group-theoretic structure of graph theory
and generalized graph theory has been studied in Refs. [25-31].
In what follows, we apply the general theory of Lie $h$-invariant
CFTs to characterize and study the $Lie$ $h$-$invariant$
$graphs$, which classify the Lie $h$-invariant level-families of
$SO(n)_{diag}$.

\bigskip\noindent
{\bf 7.2} {\bf A Review of SO(n)$_{diag}$}
\bigskip

The standard Cartesian basis of $SO(n)$ has the adjoint index
$$a=ij\,\,\,\,,\,\,\,\,1\leq i<j\leq n\eqno(7.2.1)$$
where $i$ and $j$ are vector indices of $SO(n)$, and the basic non-zero
structure constants of Cartesian $SO(n)$ are
$$f_{ij,il,jl}=-\sqrt{\tau_{n}\psi^{2}_{n}\over 2}\;\,\,,\;\,\,\tau_{n}=\left\{
\matrix{1\,\,\,,\,\,\,n\not= 3\cr 2\,\,\,,\,\,\,n=3}\right. \eqno(7.2.2)$$
where $\psi_{n}$ is the highest root of $SO(n)$ and $\tau_{n}$
is the embedding index of Cartesian $SO(n)$ in Cartesian $SO(p>n)$.
The other non-zero structure constants of this basis may be obtained from
(7.2.2) by permutation of the adjoint indices $ij$, $il$ and $jl$.

The diagonal ansatz $SO(n)_{diag}$ on $SO(n)$ has the form
$$L^{ab}=L^{ij,kl}= L_{ij}\psi_{n}^{-2} \d_{ik}\d_{jl}\eqno(7.2.3)$$
and the reduced master equation of $SO(n)_{diag}$
$$L_{ij}(1-xL_{ij})+\tau_{n}\sum_{l\not= i,j}\left[ L_{il}L_{lj}-L_{ij}(
L_{il}+L_{jl})\right]=0\eqno(7.2.4a)$$
$$c=x\sum_{i<j}L_{ij}\eqno(7.2.4b)$$
follows from (2.5), (7.2.2) and (7.2.3), where $x$ is the level of
affine $SO(n)$.
The symmetry conditions $L_{ij}=L_{ji}$, $L_{ii}=0$ are assumed in (7.2.4a).

The $2^{n\choose 2}$ level-families $L_{ij}(G_{n},x)$ of $SO(n)_{diag}$ are in
one to one correspondence with the labeled graphs $G_{n}$ of order $n$, and, in
particular, the high-level form of each level-family
$$L_{ij}(G_{n},x)={1\over x}\theta_{ij}(G_{n})+{\cal O}(x^{-2})\,\,\,,\,\,\,
\theta_{ij}(G_{n})\in\{0,1\}\eqno(7.2.5)$$
shows the adjacency matrix $\theta_{ij}(G_{n})$ of its graph $G_{n}$.
Here are some basic facts about the correspondence, which will be useful below.

1. The values $i$ or $j=1,\cdots,n$ of the $SO(n)$ vector index are the graph
   points of $G_{n}$ and the values $ij$ of the $SO(n)$ adjoint index are the
possible graph edges of $G_{n}$.

2. The graph isomorphisms
$$G'_{n}\sim G_{n}\;{\rm when} \;\theta_{ij}(G'_{n})=\theta_{\pi(i)\pi(j)}
(G_{n})\;,\;\pi\,\in\,S_{n}\,\subset\,Aut SO(n)\eqno(7.2.6)$$
are the residual  isomorphisms of $SO(n)_{diag}$
$$L_{ij}(G'_{n},x)=L_{\pi(i)\pi(j)}(G_{n},x)=\sum_{k<l}L_{kl}((\o^{-1}_{\pi})
_{kl,ij})^{2}\;\,,\;\,\o_{\pi}\in\,S_{n}\,.\eqno(7.2.7)$$
It follows that the automorphically-inequivalent
level-families of $SO(n)_{diag}$ live on the unlabeled graphs of order $n$.

3. The level-family of a graph has the symmetry of its graph,
$$L_{\pi (i)\pi(j)}(G_{n},x)=L_{ij}(G_{n},x)\,\,\,,\,\,\,\forall\pi\,\in\,
auto\,G_{n}\,.\eqno(7.2.8)$$
For each possible graph symmetry $H\subset S_{n}\subset Aut SO(n)$,
the linear relations (7.2.8)
define the consistent graph-symmetry subansatz$^{25}$ which collects the
$H$-invariant level-families of $SO(n)_{diag}$.
Historically, these were the first examples of the $H$-invariant ans\"atze
(3.1), and, as we shall see, they include the Lie $h$-invariant
subans\"atze of $SO(n)_{diag}$ as special cases  (see Section 7.4).

4. The affine-Sugawara construction on $SO(n)$ lives on the complete
graph $K_{n}$,
$$L_{SO(n)}=L(K_{n},x)\;.\eqno(7.2.9)$$

5. $SO(n)_{diag}$ contains only the subgroup constructions
$$L_{h(SO(n)_{diag})}=L(G_{h(SO(n)_{diag})},x)\eqno(7.2.10a)$$
$$G_{h(SO(n)_{diag})}=\cup_{i=1}^{N}K_{m_{i}}\eqno(7.2.10b)$$
$$h(SO(n)_{diag})=\times_{i=1}^{N} SO(m_{i})\,\,\,,\,\,\,
\sum_{i=1}^{N}m_{i}=n\eqno(7.2.10c)$$
on those subgroups $h(SO(n)_{diag})$ whose generators are a subset of the
generators $J_{ij}$ of Cartesian $SO(n)$.

6. K-conjugate level-families live on complementary graphs,
$${\tilde L}(G_{n},x)=L({\tilde G}_{n},x)\,\,\,,\;{\tilde G}_{n}=K_{n}-G_{n}
\eqno(7.2.11)$$
so that, e.g., ${\tilde L}(K_{n},x)=L({\tilde K}_{n},x)=0$, where ${\tilde
K}_{n}$ is the completely disconnected graph of order $n$.

7. The coset constructions of $SO(n)_{diag}$ live on the $g/h$ coset graphs
$$\eqalignno{&L_{SO(n)/h(SO(n)_{diag})}=L(G_{SO(n)/h(SO(n)_{diag})},x)
&(7.2.12a)\cr
&G_{SO(n)/h(SO(n)_{diag})}=K_{n}-\cup_{i=1}^{N}K_{m_{i}}\;,\;\;
\sum_{i=1}^{N}m_{i}=n &(7.2.12b)\cr}$$
which are the complete $N$-partite graphs of order $n$.

8. The affine-Sugawara nests (2.8) live on the affine-Sugawara nested
   graphs, for example,
$$L_{SO(n)/SO(m_{1})/\cdots/SO(m_{d})}=L(G_{SO(n)/SO(m_{1})/\cdots/SO(m_{d})},
x)\eqno(7.2.13a)$$
$$G_{SO(n)/SO(m_{1})/\cdots/SO(m_{d})}=K_{n}-(K_{m_{1}}-(K_{m_{2}}-(
\cdots-K_{m_{d}}))\cdots)\;. \eqno(7.2.13b)$$
These are the graphs of the ``old" or standard rational conformal field
theories in $SO(n)_{diag}$, and the more general affine-Virasoro nested
graphs are formed from these by the replacement $K_{m_{d}}\rightarrow
G_{m_{d}}$, where $G_{m_{d}}$ is any graph of order $m_{d}$.

9. The self K-conjugate level-families
$${\tilde L}(G_{n},x)=L({\tilde G}_{n},x)=\o L(G_{n},x)\o^{-1}\;,\;\o\,\in\,
S_{n}\eqno(7.2.14)$$
live on the   self-complementary graphs ${\tilde G}_{n}\sim G_{n}$.
\bigskip

\noindent
See Ref. [28] for further discussion of the Lie group-theoretic structure of
graph theory including the Lie-algebraic form of the edge-adjacency matrix and
the isomorphism groups of graph theory and the generalized graph theories on
Lie
$g$.

\bigskip\noindent
{\bf 7.3} {\bf Lie {\bfi h}-Invariant Subans\"atze in SO(n)$_{diag}$}
\bigskip

According to (4.1) and (7.2.3), the $h(SO(n)_{diag})$-invariant subansatz
of $SO(n)_{diag}$ is
$$   (L_{ij}-L_{kl})\sum_{r<s}f_{ij,kl,rs}\psi_{rs}=0\eqno(7.3.1)$$
where $\psi_{rs}$ parametrizes $h(SO(n)_{diag})$ in the vicinity of the origin.
In fact, the sum in (7.3.1) may be dropped, and we may write
$$A_{SO(n)_{diag}}(h(SO(n)_{diag}))\,\,:\,\,(L_{ij}-L_{kl})f_{ij,kl,rs}=0\,\,\,,
\,\,\,\forall\,rs\,\in\,h(SO(n)_{diag})\eqno(7.3.2)$$
because the generators of $h(SO(n)_{diag})$ are a subset of the generators of
Cartesian $SO(n)$.

As an example, we consider the $SO(m)$-invariant subansatz
in $SO(n)_{diag}$.
Decompose the vector indices of $SO(n)$ as
$$i=(\mu,I)\,\,\,,\,\,\,i=1,\cdots,m\,\,\,,\,\,\,I=m+1,\cdots,n\eqno(7.3.3)$$
so that Greek letters are vector indices of $SO(m)$.
Then the non-zero coefficients of the ansatz
$$A_{SO(n)_{diag}}(SO(m))\;\,\,:\;\;\;\,\,L_{\mu\nu}=L\,\,,\,\,L_{\mu I}=
L_{I}\,\,,\,\,L_{IJ}\eqno(7.3.4)$$
are obtained from (7.3.2) with the structure constants (7.2.2).
The reduced VME of the $SO(m)$-invariant subansatz
$$\tau_{n}^{-1}L(1-xL)-(m-2)L^{2}+\sum_{I}L_{I}(L_{I}-2L)=0\eqno(7.3.5a)$$
$$\tau_{n}^{-1}L_{I}(1-xL_{I})-(m-1)L_{I}^{2}+\sum_{J\not= I}\left[ L_{IJ}(
L_{J}-L_{I})-L_{I}L_{J}\right]=0\eqno(7.3.5b)$$
$$\eqalignno{\tau_{n}^{-1}L_{IJ}(1-xL_{IJ})+&m(L_{I}L_{J}-L_{IJ}(L_{I}+L_{J}))
\cr
&+\sum_{K\not= I,J}\left[L_{IK}L_{KJ}-L_{IJ}(L_{IK}+L_{JK})\right]=0&(7.3.5c)
\cr}$$
follows from (7.2.4).
Any other decomposition $i=(\mu,I)$ with dim$\,\mu=m$ gives an automorphically
equivalent copy of $A_{SO(n)_{diag}}(SO(m))$.

Similarly, the $(SO(m_{1})\times SO(m_{2}))$-invariant subansatz of $SO(n)
_{diag}$
$$L_{\mu_{1}\nu_{1}}=L_{1}\,\,\,,\,\,\,L_{\mu_{2}\nu_{2}}=L_{2}\;,\;
L_{\mu_{1}\mu_{2}}=L\eqno(7.3.6a)$$
$$L_{\mu_{1}I}=L_{I,1}\;,\;L_{\mu_{2}I}=L_{I,2}\;,\;L_{IJ}\eqno(7.3.6b)$$
is obtained with $i=(\mu_{1},\mu_{2},I)$, where $\mu_{1}$ (or $\nu_{1}$) and
$\mu_{2}$ (or $\nu_{2}$) are
vector indices of $SO(m_{1})$ and $SO(m_{2})$ respectively.

\bigskip\noindent
{\bf 7.4} {\bf The Lie {\bfi h}-Invariant Graphs}
\bigskip

The Lie $h$-invariant level-families of $SO(n)_{diag}$ are in one to one
correspondence
with the $Lie$ $h$-$invariant$ $graphs$, whose characterization
may be obtained by taking
the high-level limit (7.2.5) of the Lie $h$-invariant
level-families (7.3.2):\hfill\break
\noindent
A graph $G_{n}$ of order $n$, with adjacency matrix $\theta_{ij}(G_{n})$,
is (at least) Lie $h(SO(n)_{diag})$-invariant when
$$(\theta_{ij}(G_{n})-\theta_{kl}(G_{n}))f_{ij,kl,rs}=0\,\,\,,\,\,\,\forall\,
rs\,\in\,h(SO(n)_{diag})\eqno(7.4.1)$$
where the subgroups $h(SO(n)_{diag})$ are characterized in (7.2.10) and
$f_{ij,kl,rs}$ are the structure constants (7.2.2) of Cartesian $SO(n)$.
The equivalent Lie group characterization
$$\Theta(G_{n})=\o \Theta(G_{n})\o^{-1}\eqno(7.4.2a)$$
$$\Theta_{ij,kl}\equiv \delta_{ik}\delta_{jl}\theta_{ij}(G_{n})\,\,\,,\,\,\,
\forall\,\o_{ij,kl}\,\in\,h(SO(n)_{diag})\eqno(7.4.2b)$$
also follows with (7.2.3) as the high-level limit of (3.1).

The forms (7.4.1) and (7.4.2) are the Lie-algebraic and Lie group-theoretic
characterization of the Lie $h$-invariant graphs.
The Lie characterization of these graphs is another facet of the underlying Lie
group-theoretic structure of graph theory and generalized graph theory on Lie
$g$ \cite{28}.

To obtain a visual characterization of the Lie $h$-invariant graphs ,
we must solve (7.4.1) or (7.4.2).
Consider first the $SO(m)$-invariant graphs of order $n$,
and let $i=(\mu,I)$ where $\mu$ (the vector index of $SO(m)$)
runs over any set of $m$ graph points while $I$ runs over the
remaining $n-m$ points.
The solution of (7.4.1) for $rs=\mu\nu\,\in\,SO(m)$
$$\matrix{\hbox{{\it SO(m)}-invariant}\cr
\hbox{graphs of}\cr \hbox{order}\;n\cr}\;\;\;\;\;\left\{\;\;\;\;\;\matrix{
\theta_{\mu\nu}(G_{n})\;\hbox{is independent of}\;\mu,\nu\cr
\theta_{\mu I}(G_{n})\;\hbox{is independent of}\;\mu\cr
\theta_{IJ}(G_{n})\;{\rm arbitrary}\cr}\right.\eqno(7.4.3)$$
follows easily with the structure constants (7.2.2), or as the
high-level limit (7.2.5) of the
$SO(m)$-invariant conformal level-families in (7.3.4).

The $SO(m)$-invariant graphs of order $n$ are shown schematically in Fig. 2,
which distinguishes two cases,
$$G_{n-m}\oplus_{p}{\tilde K}_{m}\;\;\;(\theta_{\mu\nu}=0)\eqno(7.4.4a)$$
$$G_{n-m}\oplus_{p}K_{m}\;\;\;(\theta_{\mu\nu}=1)\;.
\eqno(7.4.4b)$$
Here, $G_{n-m}$ is any graph of order $n-m$, and the $partial$ $join$
$\oplus_{p}$ is defined to connect $p\leq n-m$ points of $G_{n-m}$ to all
points in ${\tilde K}_{m}$ or $K_{m}$.

It is also clear from Fig. 2 that the $SO(m)$-invariant graphs are those
with a $graph$-$local$ discrete symmetry $S_{m}$, which, by definition,
acts only on a fixed set of $m$ points, without permutation of the remaining
$n-m$ points of the graph.
Conversely, any graph with a graph-local $S_{m}$ symmetry is also
$SO(m)$-invariant.

Similarly, to characterize the $(SO(m_{1})\times SO(m_{2}))$-invariant graphs
of order $n$, decompose the $SO(n)$ vector index as $i=(\mu_{1},\mu_{2},I)$,
where $\mu_{1}$ (or $\nu_{1}$) and $\mu_{2}$ (or $\nu_{2}$) run over
distinct sets of $m_{1}$ and $m_{2}$ graph points respectively
and $I$ runs over the remaining $n-m_{1}-m_{2}$ points.
Then, the characterization of the
$(SO(m_{1})\times SO(m_{2}))$-invariant graphs
$$\theta_{\mu_{1}\nu_{1}}(G_{n})\;,\;\theta_{\mu_{2}\nu_{2}}(G_{n})\;,\;
\theta_{\mu_{1}\mu_{2}}(G_{n})\;{\rm are}\;{\rm  independent}\;{\rm  of}\;
{\rm their}\;{\rm labels}\eqno(7.4.5a)$$
$$\theta_{\mu_{1}I}(G_{n})\;,\;\theta_{\mu_{2}I}(G_{n})\;{\rm are}\;
{\rm independent}\;{\rm of}\;\mu_{1}\,,\,\mu_{2}\eqno(7.4.5b)$$
$$\theta_{IJ}(G_{n})\;{\rm arbitrary}\eqno(7.4.5c)$$
is obtained as the high-level limit of the
$(SO(m_{1})\times SO(m_{2}))$-invariant level-families in (7.3.6).
These graphs are shown schematically in Fig. 3, where $G_{n-m_{1}-m_{2}}$ is
a general graph of order $n-m_{1}-m_{2}$.
As above, the subgroup components ($K_{m_{1}}$ or ${\tilde K}_{m_{1}}$) and
($K_{m_{2}}$ or ${\tilde K}_{m_{2}}$) are connected to $G_{n-m_{1}-m_{2}}$ by
the independent partial joins $\oplus_{p_{1}}$ and $\oplus_{p_{2}}$;
the two cases shown in the figure correspond to $\theta_{\mu_{1}\mu_{2}}=0$ or
$1$, that is, the connection of all or none of the points of
(K$_{m_{1}}$ or ${\tilde {\rm K}}_{m_{1}}$) to the points of (K$_{m_{2}}$
or ${\tilde {\rm K}}_{m_{2}}$).

The general $h(SO(n)_{diag})$-invariant graph is constructed as follows:

1. Choose the subgroup components $h_{i}$=$({\tilde K}_{m_{i}}$ or $K_{m_{i}})$
   for each $SO(m_{i})$ in $h(SO(n)_{diag})$$=\cup_{i=1}^{N}SO(m_{i})$,
   $m_{i}\geq 2$, $\sum_{i=1}^{N}m_{i}\leq n$.

2. Choose a set of independent partial joins $\{\oplus_{p_{i}}$, $p_{i}\leq
   n-\sum_{i=1}^{N}m_{i}\}$ to connect each subgroup component $h_{i}$ to any
   graph of order $n-\sum_{i=1}^{N}m_{i}$.

3. Choose to join or not to join each pair of subgroup components $h_{i}$ and
   $h_{j}$.

\noindent
Moreover, a graph-local discrete symmetry $S_{m_{i}}$ is associated to each
${\tilde K}_{m_{i}}$ or $K_{m_{i}}$ in the general Lie $h$-invariant graph.
In particular, it was noted in Ref. [25] that every affine-Sugawara nested
graph
has at least a $Z_{2}$ symmetry, and we have checked that this $Z_{2}$ is the
graph-local $S_{2}$ associated to the minimal Lie symmetry $SO(2)$ of any
affine-Sugawara nest.
The conformal field-theoretic interpretation of ${\tilde K}_{m}$ or $K_{m}$
is given in Section 8.

\bigskip\noindent
{\bf 7.5} {\bf On the Number of Lie {\bfi h}-Invariant Graphs}
\bigskip

In this section, we discuss the (unlabeled or labelled) graph hierarchy:
$${\rm graphs}\phantom{WWWWWWWWWWWWWWWWWWWWW}$$
$$>>\;{\rm graphs}\;{\rm with}\;{\rm symmetry}\phantom{WWWWWWWW}$$
$$>>\;{\rm Lie}\;h{\rm -invariant}\;{\rm graphs}\phantom{W}$$
$$\phantom{WWWWWWWWWWWW}>>\;{\rm affine-Sugawara}\;{\rm nested}\;{\rm graphs}
\eqno(7.5.1)$$
which tells us, as in Section 4, that the Lie $h$-invariant level-families,
though copious in $SO(n)_{diag}$, are not generic.
The last inequality also shows that the generic Lie $h$-invariant
level-family is a set of new (generically unitary and irrational) CFTs.

It is well known that the generic graph has no symmetry, and it is clear that
the Lie $h$-invariant graphs, having a special symmetry, are not generic among
the graphs with symmetry.

Bounds on the number of Lie $h$-invariant graphs are obtained as follows.
The number of labelled $SO(m)$-invariant graphs in the subansatz
$A_{SO(n)_{diag}}(SO(m))$ is
$$|A_{SO(n)_{diag}}(SO(m))|=2^{1+{(n-m)(n-m+1)\over 2}}\eqno(7.5.2)$$
where the exponent is the number of unknowns in the subansatz (see Eq.
(7.3.4)).
The number (7.5.2) is a lower bound on the number of labelled $SO(m)$-invariant
graphs of order $n$ because the subansatz corresponds to the choice of a
fixed $SO(m)$, that is, a fixed set of graph points, say $\mu=1,\cdots,m$.
Moreover, (7.5.2) is an upper bound on the number of unlabeled $SO(m)
$-invariant graphs of order $n$ because the subansatz contains at least one
labelled representative of the equivalence class of each $SO(m)$-invariant
graph.
We conclude that the number of inequivalent $SO(m)$-invariant level-families
of $SO(n)_{diag}$ satisfies
$${1\over n!}2^{1+{(n-m)(n-m+1)\over 2}} \leq N\left(
\matrix{\hbox{unlabelled}\;SO(m)\hbox{-invariant}\cr
\hbox{graphs of order}\;n\cr}\right)\leq2^{1+{(n-m)(n-m+1)\over 2}}
\eqno(7.5.3)$$
because no graph of order $n$ is isomorphic to more than $n!$ other graphs.
Moreover, the total number of inequivalent Lie $h$-invariant constructions in
$SO(n)_{diag}$ satisfies
$${1\over n!}2^{n^{2}-3n+4\over 2}\leq N\left(\matrix{\hbox{unlabelled Lie}\;
h\;\hbox{-invariant}\cr
\hbox{graphs of order}\;n\cr}\right)\leq 2^{n^{2}-3n+4\over 2}\eqno(7.5.4)$$
because every Lie $h$-invariant graph has at least an $SO(2)$ symmetry.

Comparison of these inequalities to the total number of ``old" Lie
$h$-invariant level-families
in $SO(n)_{diag}$ \cite{25}
$$N\left(\matrix{\hbox{unlabelled affine-Sugawara}\cr
\hbox{nested graphs of order}\;n\cr}\right) \leq {1\over 2n-1}{2n\choose n}=
{\cal O}(e^{2n\log 2})\,.\eqno(7.5.5)$$
shows that the generic Lie $h$-invariant level-family in $SO(n)_{diag}$ is a
set of new CFTs.

Some precise counting of Lie $h$-invariant graphs on small manifolds is given
in Section 10.1.

\bigskip\noindent
{\large\bf 8.} {\large\bf Graphical Identification of (0,0) and (1,0)
Operators}
\bigskip

Recall from Section 5 that the symmetry of Lie $h$-invariant CFTs is
$h=h_{0}\oplus h_{1}$, where $h_{0}$ and $h_{1}$ are the affine and global
components of $h$ respectively.

To understand this in the graphs, consider the conformal weights
$\Delta_{ij}(G_{n},x)$ of the one-current states $J^{(-1)}_{ij}|0\rangle$
of the level-family $L(G_{n},x)$ of each graph (see Appendix A):
In $SO(n)_{diag}$, the matrix ${M_{a}}^{b}(L)$ is diagonal
$${M_{ij}}^{kl}(G_{n},x)=\Delta_{ij}(G_{n},x)\delta_{ik}\delta_{jl}\eqno(8.1)$$
so that a conformal weight is associated to each pair of points $ij$ in
$G_{n}$, and we obtain
$$\Delta_{ij}(G_{n},x)=xL_{ij}(G_{n},x)+{1\over 2}\sum_{l\not= i,j}(
L_{il}(G_{n},x)+L_{jl}(G_{n},x))\eqno(8.2a)$$
$$=\theta_{ij}(G_{n})+{\cal O}(x^{-2})\;.\eqno(8.2b)$$
For most $ij$ in most graphs, the conformal weights in (8.2a) will be
irrational at finite level.
For the Lie $h$-invariant level-families, on the other hand, we recall the
exact result of Section 5,
$$\Delta_{ij}(G_{n},x)=\left\{\matrix{0\;\;,\;\;ij\,\in\,h_{0}\cr
1\;\;,\;\;ij\,\in\,h_{1}\cr}\right.\eqno(8.3)$$
which shows that the higher-order corrections in (8.2) will vanish when
$ij\in h$.
Comparing (8.2) and (8.3) with (7.4.4), we learn that
$$J_{ij}\;{\rm is}\; {\rm a}\; (0,0)\;{\rm operator}\;{\rm of}\;L(G_{n-m}
\oplus_{p}{\tilde K}_{m})\;{\rm when}\;i,j\in{\tilde K}_{m}\eqno(8.4a)$$
$$J_{ij}\;{\rm is}\; {\rm a}\; (1,0)\;{\rm operator}\;{\rm of}\;
L(G_{n-m}\oplus_{p} K_{m})\;{\rm when}\;i,j\in K_{m}\eqno(8.4b)$$
More generally, it follows that the Lie symmetry $SO(m_{i})\subset h(SO(n)
_{diag})$ is affine for all
subgroup components ${\tilde K}_{m_{i}}$ and global for all
subgroup components  $K_{m_{i}}$.
In what follows, we refer to the set of graphs with any affine $h$-invariance
as the $\hh$-$invariant$ $graphs$.

\bigskip\noindent
{\large\bf 9.} {\large\bf The Lie {\bfi h}-Invariant Graph Multiplets}
\bigskip

The $Lie$ $h$-$invariant$ $graph$ $multiplets$ are the Lie $h$-invariant
conformal multiplets of Section 6, restricted to the ansatz $SO(n)_{diag}$.

As an example, begin with the ${\widehat SO(m)}$-invariant graph $G_{n-m}
\oplus_{p}{\tilde K}_{m}$, and consider the $SO(m)$-invariant graph quartet
on $SO(n)$
$$  \matrix{ G_{n-m}\oplus_{p}{\tilde K}_{m} & \mapright{+SO(m)} &
G_{n-m}\oplus_{p}K_{m} \cr
\mapupdown{{\rm K}_{SO(n)}} & &
\mapupdown{{\rm K}_{SO(n)}} \cr
{\tilde G}_{n-m}\oplus_{n-m-p}K_{m} & \mapleft{+SO(m)}
& {\tilde G}_{n-m}\oplus_{n-m-p}{\tilde K}_{m} } \eqno(9.1)$$
which corresponds to (6.3a).
The graph moves here are
$$+SO(m)\equiv\hbox{add the edges of }K_{m}\,=\,+L_{SO(m)}\eqno(9.2a)$$
$${\rm K}_{SO(n)}=\hbox{complementarity in the graphs of order }n\;.
\eqno(9.2b)$$
The K$_{g/h}$-conjugation in (6.6)

$$  \matrix{ G_{n-m}\oplus_{p}{\tilde K}_{m} & &
G_{n-m}\oplus_{p}K_{m} \cr
&\mapdiagr{{\rm K}_{g/h}} & \cr
{\tilde G}_{n-m}\oplus_{n-m-p}K_{m} &
& {\tilde G}_{n-m}\oplus_{n-m-p}{\tilde K}_{m} } \eqno(9.3)$$
is realized in these graphs as the K$_{SO(n)/SO(m)}$-complementarity
$$\eqalignno{{\rm K}_{SO(n)/SO(m)}\;:\;\;&{\tilde G}_{n-m}\oplus_{n-m-p}{\tilde
K}_{m}=G_{SO(n)/SO(m)}-G_{n-m}\oplus_{p}{\tilde K}_{m}\phantom{WW}&(9.4a)\cr
&G_{SO(n)/SO(m)}=K_{n}-K_{m}=K_{n-m}+{\tilde K}_{m}&(9.4b)\cr}$$
through the coset graph $G_{SO(n)/SO(m)}$.
The K$_{g+h}$-complementarity along the other diagonal of (9.1)
is the composite move K$_{SO(n)}$ followed by $+SO(m)$.

More generally, K$_{g/h}$-$complementarity$ through the general $g/h$ coset
graph
$$\eqalignno{{\rm K}_{SO(n)/h(SO(n)_{diag})}:\;\;&({\tilde G}_{n})_{SO(n)
/(h(SO(n)_{diag}))}\equiv G_{SO(n)/(h(SO(n)_{diag}))}-G_{n}\phantom{WWW}&(9.5a)
\cr &G_{SO(n)/(h(SO(n)_{diag}))}= K_{n}-\cup_{i=1}^{N}K_{m_{i}}\;,\;\;
\sum_{i=1}^{N}m_{i}=n&(9.5b)\cr}$$
satisfies K$_{g/h}^{2}=1$ on the space of $\hh(SO(n)_{diag})$-invariant graphs.

The self K$_{g/h}$-conjugate level-families of $SO(n)_{diag}$ live with
$c={1\over 2}c_{g/h}$ on the $self$ K$_{g/h}$-$complementary$ $graphs$, which
satisfy
$$({\tilde G}_{n})_{SO(n)/h(SO(n)_{diag})}\sim G_{n}\;.\eqno(9.6)$$
Examples of these graphs are given in Section 10, and the exact self
K$_{g/h}$-conjugate level-families in $SO(6)_{diag}$ are obtained in Section
11.4.
When a Lie $h$-invariant quartet contains a self K$_{g/h}$-complementary graph,
we know from Section 6 that the graphs along the other diagonal are also
isomorphic (self K$_{g+h}$-complementary), with level-families whose central
charge is $c=(c_{g}+c_{h})/2$.

The higher Lie $h$-invariant graph multiplets can be studied from their
corresponding conformal multiplets, viz. Fig. 1: In particular, a
$2^{d+1}$-plet of Lie $h_{d}$-invariant graphs is associated to the general
subgroup nest $SO(n)\supset h_{1}\supset\cdots\supset h_{d}$, $h_{i}\in h(SO(n)
_{diag})$.
These ($d+1$)-dimensional cubes may be generated from any $\hh_{1}$-invariant
graph by the graph moves K$_{SO(n)}$ and $+h_{i}$, $i=1,\cdots,d$, and the
multiplets show a broad variety of generalized complementarities, including
K$_{SO(n)/h_{1}/\cdots/h_{d}}$-$complementarity$ through
the general affine-Sugawara nested graph $G_{SO(n)/h_{1}\cdots/h_{d}}$.
Because self-complementary and self K$_{g/h}$-complementary graphs exist, we
conjecture that generalized self-complementary graphs exist for all the
generalized complementarities.
Examples of Lie $h$-invariant graph octets are given in the next section.

\bigskip\noindent
{\large\bf 10.} {\large\bf Examples of Lie {\bfi h}-Invariant Graphs}

\bigskip
\noindent
{\bf 10.1.} {\bf Counting on Small Manifolds}
\bigskip

On $SO(2)$ and $SO(3)$, all graphs are affine-Sugawara nested graphs,
and hence at least $SO(2)$-invariant.

On $SO(4)$, 10 of the 11 unlabeled graphs are Lie $h$-invariant.
These 10 graphs are arranged in Fig. 4 as an $SO(2)$-invariant octet on
$SO(4)\supset SO(3)\supset SO(2)$ and an ($SO(2) \times SO(2)$)-invariant
quartet on $SO(4)\supset SO(2)\times SO(2)$.
The graphs on the top face of the cube form an $SO(3)$-invariant quartet.
The 11th graph of $SO(4)$, with no Lie $h$-symmetry, is the self-complementary
path graph of order 4.
More generally, we have found no self-complementary graphs which
are also Lie $h$-invariant.

On $SO(5)$, we find that 28 of the 34 unlabeled graphs are Lie $h$-invariant.
These include 24 affine-Sugawara nested graphs and an $SO(2)$-invariant
quartet, shown in Fig. 5, whose conformal level-families are new.
The coset graph $G_{SO(5)/SO(2)}=K_{5}-K_{2}$, through which
K$_{g/h}$-complementarity acts in this case, is the sum of the edges of the
K$_{g/h}$-complementary graphs in the quartet.

Among the 156 unlabeled graphs of $SO(6)$, we find 120 Lie $h$-invariant
graphs, including 66 affine-Sugawara nested graphs and 54 Lie $h$-invariant
graphs whose level-families are new.
Among the graphs of the new level-families, 42 are new
irreducible\footnote{The new irreducible graphs \cite{25} are those which
cannot be constructed by affine-Virasoro nesting from smaller manifolds.}
graphs, which can be arranged as follows:

a) 1 $SO(3)$-invariant quartet in an $SO(2)$-invariant octet on $SO(6)\supset
   SO(3)\supset SO(2)$.

b) 2 $(SO(2)\times SO(2))$-invariant octets on $SO(6)\supset SO(2)\times SO(2)
   \supset SO(2)$. One of these octets has 2 pairs of isomorphic graphs (see
   Section 11.2).

c) 2 $SO(2)$-invariant quartets on $SO(6)\supset SO(2)$ which contain
   self K$_{SO(6)/SO(2)}$-complementary graphs (see Section 10.2 and 11.4).

d) 4 other $SO(2)$-invariant quartets on $SO(6)\supset SO(2)$ (see Section
   11.3).

As an example on $SO(n)$, we mention the $SO(n-3)$-invariant graphs of Fig. 6,
which are the new irreducible $\hh$-invariant graphs with the largest Lie
symmetry on each manifold.

\bigskip\noindent
{\bf 10.2.} {\bf Self K$_{g/h}$-Complementary Graphs}
\bigskip

Fig. 7 shows the 2 $SO(2)$-invariant quartets on $SO(6)$ which contain self
K$_{g/h}$-complementary graphs, and the graphical form of K$_{g/h}$
is shown under each quartet.
According to the general theory, both of these self K$_{g/h}$-conjugate
level-families have central charge
$$c={1\over 2}c_{SO(6)/SO(2)}={7x-2\over x+4}\eqno(10.2.1)$$
while the self K$_{g+h}$-conjugate level-families on the other diagonal of the
quartets have
$c=(c_{SO(6)}+c_{SO(2)})/2=2(4x+1)/(x+4)$.

We consider self K$_{g/h}$-complementary graphs in further detail for the case
$g/h=SO(n)/SO(m)$.
According to (9.3) and (9.6), the self K$_{SO(n)/SO(m)}$-complementary graphs
satisfy
$$G_{n-m}\oplus_{p} {\tilde K}_{m}\,\sim\,{\tilde
G}_{n-m}\oplus_{n-m-p}K_{m}\eqno(10.2.2)$$
so the generic self K$_{SO(n)/SO(m)}$-complementary graph has the structure
$$G_{4q}\oplus_{2q}{\tilde K}_{m=n-4q}\;\;,\;\;{\tilde G}_{4q}\sim G_{4q}\eqno(
10.2.3)$$
which is a half-partial join of ${\tilde K}_{m}$ to a self-complementary
graph of order $4q$.
The structure (10.2.3) predicts self K$_{g/h}$-complementary graphs on $SO(n)
\supset SO(n-4q)$, $n\geq 6$, and the central charges of the corresponding
self K$_{SO(n)/SO(n-4q)}$-conjugate level-families of these graphs are
$$c={1\over 2}c_{SO(n)/SO(n-4q)}\;\;,\;\;n\geq 6\;.\eqno(10.2.4)$$
The examples in Fig. 7 on $SO(6)\supset SO(2)$ are constructed with $q=1$
on the single self-complementary graph $G_{4}=P_{4}$ of order 4, and these
examples are the only self
K$_{g/h}$-complementary graphs on $SO(6)$.
Fig. 8 shows the entire $q=1$ series of self K$_{SO(n)/SO(n-4)}$-complementary
graphs (constructed on $P_{4}$), and, following (10.2.3), we have also
constructed 36 distinct series of self K$_{SO(n)/SO(n-8)}$-complementary
graphs on the 10 self-complementary graphs of $SO(8)$.

\bigskip\noindent
{\large\bf 11.} {\large\bf Exact Solutions on Lie {\bfi h}-Invariant Graphs}

\bigskip
\noindent
{\bf 11.1.} {\bf Small Subans\"atze with New Level-Families}
\bigskip

Smaller ans\"atze are generally more amenable to exact solution, and these
occur with higher symmetry on smaller manifolds.
The Lie $h$-invariant subans\"atze in $SO(5)_{diag}$ and $SO(6)_{diag}$
which contain new level-families are

$$SO(5)[d(SO(2)),7]\equiv A_{SO(5)_{diag}}(SO(2))$$
$$SO(6)[d(SO(3)),7]\equiv A_{SO(6)_{diag}}(SO(3))$$
$$SO(6)[d(SO(2)\times SO(2)),8]\equiv A_{SO(6)_{diag}}(SO(2)\times SO(2))$$
$$SO(6)[d(SO(2)),11]\equiv A_{SO(6)_{diag}}(SO(2))\eqno(11.1.1)$$
where we have introduced the nomenclature of Ref. [25]
to show the number of unknowns in the subansatz.

\bigskip\noindent
{\bf 11.2.} {\bf An (SO(2)$\times$SO(2))-Invariant Octet in SO(6)$_{diag}$}
\bigskip

Because it has a subansatz of higher symmetry, we focus first on the
8-dimensional $(SO(2)\times SO(2))$-invariant subansatz of $SO(6)_{diag}$,
$$L_{15}=L_{25}\;,\;L_{16}=L_{26}\;,\;L_{35}=L_{45}\;,\;L_{36}=L_{46}$$
$$L_{13}=L_{23}=L_{14}=L_{24}$$
$$L_{12}\;,\,\;L_{34}\;,\;L_{56}\eqno(11.2.1)$$
where we have chosen $1,2\in SO(2)$ and $3,4\in SO(2)'$.
This subansatz contains 14 new irreducible $(SO(2)\times SO(2))$-invariant
level-families, whose graphs form the 2 $(SO(2)\times SO(2))$-invariant octets
shown in Fig. 9.

All 256 labelled graphs of the subansatz have the graph-local discrete symmetry
$$S_{2}\times S_{2}\;\;:\;\;1\leftrightarrow 2 \;\;\;\hbox{and/or}\;\;\;3
\leftrightarrow 4\eqno(11.2.2)$$
associated to their $(SO(2)\times SO(2))$ invariance, but those graphs for
which both $SO(2)$s are affine (or both global) have the higher symmetry
$SO(2)\times SO(2)\times Z_{2}$, where the discrete symmetry
$$Z_{2}\;\;\;:\;\;\; 1\leftrightarrow 3\;\;,\;\;2\leftrightarrow 4\;\;,\;\;
5\leftrightarrow 6\eqno(11.2.3)$$
is the exchange symmetry of the two $SO(2)$ subgroups.
Among the 14 new irreducible graphs, only 4 of the graphs in the right octet of
Fig. 9 have this symmetry, and we have redrawn two of them in Fig. 10 to show
the $Z_{2}$ in (11.2.3) as a left-right symmetry.
The remaining 2 new irreducible ($SO(2)\times SO(2)\times Z_{2}$)-invariant
graphs are the complements of those in the figure.

The ($SO(2)\times SO(2)\times Z_{2}$)-invariant graphs of order 6 are collected
in the 5-dimensional subansatz
$$L_{15}=L_{25}=L_{36}=L_{46}$$
$$L_{16}=L_{26}=L_{35}=L_{45}$$
$$L_{13}=L_{23}=L_{14}=L_{24}$$
$$L_{12}=L_{34}\;,\;L_{56}\eqno(11.2.4)$$
which follows from (11.2.1) by imposing the symmetry (11.2.3) in the form
$L_{\pi (i)\pi (j)}$=$L_{ij}$.
In what follows, we refer to this subansatz as $SO(6)[d(SO(2)\times SO(2)),5]$,
and we have named the level-families of the first two graphs in Fig. 10
accordingly.
The reduced master equation of $SO(6)[d(SO(2)\times SO(2)),5]$
$$L_{15}(1-xL_{15})=L_{15}(L_{15}+3L_{16}+2L_{13}+L_{12}+L_{56} )
-L_{12}L_{15}-L_{16}L_{56}-2L_{16}L_{13}$$
$$L_{16}(1-xL_{16})=L_{16}(L_{16}+3L_{15}+2L_{13}+L_{12}+L_{56} )
-L_{12}L_{16}-L_{15}L_{56}-2L_{15}L_{13}$$
$$L_{13}(1-xL_{13})=2L_{13}(L_{13}+L_{12}+L_{15}+L_{16})
-2L_{12}L_{13}-2L_{15}L_{16}$$
$$L_{12}(1-xL_{12})=2L_{12}(L_{15}+L_{16}+2L_{13})-L_{15}^2-L_{16}^2-2L_{13}^2$$
$$L_{56}(1-xL_{56})=4L_{56}(L_{15}+L_{16})-4L_{15}L_{16}$$
$$c=x(4(L_{13}+L_{15}+L_{16})+2L_{12}+L_{56})\eqno(11.2.5)$$
follows from (7.2.4) and (11.2.4).

Except for possible sporadic solutions, we have solved the 5-dimensional
subansatz completely.
The new ($SO(2)\times SO(2)\times Z_{2}$)-invariant level-families are
$$L_{15}={(1+\eta R)\over 2(x+4)} +{\eta \s  \over 2}
{\sqrt{x^2-4x-4} \over x-2} R $$
$$L_{16}={(1+\eta R)\over 2(x+4)} -{\eta  \s  \over 2}
{\sqrt{x^2-4x-4} \over x-2} R$$
$$L_{13}=\frac{1}{2(x+4)}\left(1-\eta {x^2-x-14 \over x-2} R\right)$$
$$L_{12}={\eta\xi \over 2x}+\frac{1}{2(x+4)}\left(1+2\eta {x^2-2x-12 \over
x(x-2)} \right)$$
$$ L_{56} =\frac{1}{2(x+4)}\left(1+\eta {x^2-x-26 \over x-2} R\right)$$
$$c=\frac{1}{2(x+4)}\left(15x-\eta(3x^2-9x-24)R\right) +\xi \eta$$
$$R\equiv \sqrt{{ (x-2) \over (x+1) (x^2-x-18)}}\eqno(11.2.6)$$
where $\eta=\pm 1$ is K-conjugation on $SO(6)$, $\sigma =\pm 1$ labels
automorphically equivalent copies and $\xi=\pm 1$ are inequivalent
level-families.

The high-level limit of the level-families (11.2.6) establishes the
correspondence
$$\eqalignno{SO(6)^{\#}[d(SO(2)\times SO(2)),5]_{1}\;\;&:\;\; \eta=-\xi=1&(11.
2.7a)\cr
SO(6)^{\#}[d(SO(2)\times SO(2)),5]_{2}\;\;&:\;\; \eta=\xi=1&(11.2.7b)\cr}$$
with the first two graphs in Fig. 10, where $\sigma =1$ is required to obtain
the labelling shown.
In particular, the level-family $SO(6)^{\#}[d(SO(2)\times SO(2)),5]_{1}$ is
$({\widehat SO(2)}\times {\widehat SO(2)}\times Z_{2})$-invariant (with (0,0)
operators $J_{12}$ and $J_{34}$) and the globally-invariant
level-family $SO(6)^{\#}[d(SO(2)\times SO(2)),5]_{2}$ can be obtained from it
by the move $+(SO(2)\times SO(2)) $=\break
$+L_{SO(2)\times SO(2)}$, as in Fig. 9.

As a bonus, we obtain the $({\widehat SO(2)}\times SO(2))$-invariant conformal
level-family
$$L(SO(6)^{\#}[d(SO(2)\times SO(2)),8]_{1})=L(SO(6)^{\#}[d(SO(2)\times SO(2)),
5]_{1}+L_{SO(2)}\eqno(11.2.8)$$
where $SO(2)$ is either $1,2$ or $3,4$.
This level-family is the third graph in Fig. 10, and, with K-conjugation,
we have obtained all the level-families of the right
$(SO(2)\times SO(2))$-invariant octet in Fig. 9.

The level-families of this octet have generically irrational central charge
and they are unitary for integer level $x\geq 5$.
The lowest unitary irrational central charge in the octet is found at level
6
$$c(SO(6)_{6}[d,5]^{\#}_{1})={7\over 2}\left(1-{3\over 7\sqrt{21}}\right)\simeq
3.1727\eqno(11.2.9)$$
and, more generally, all the central charges in the octet increase
monotonically with the level toward the integers $c_{0}=$dim$E(G)$.

\bigskip\noindent
{\bf 11.3.} {\bf An SO(2)-Invariant Quartet in SO(6)$_{diag}$}
\bigskip

The 11-dimensional subansatz $SO(6)[d(SO(2)),11]$ in $SO(6)_{diag}$ is
$$ L_{1i}=L_{2i},\;\;\;3 \leq i \leq 6 $$
$$ L_{12}, \;\;\;L_{ij},\;\;\; 3 \leq i < j \leq 6 \eqno(11.3.1)$$
where we have chosen $1,2\in SO(2)$.
This subansatz contains representatives of all the Lie $h$-invariant
level-families of $SO(6)_{diag}$

All $2^{11}$ labelled graphs of the subansatz have the graph-local discrete
symmetry
$$ S_2\;\;:\;\;1 \leftrightarrow 2 \eqno(11.3.2)$$
associated to their $SO(2)$-invariance, but we will consider only those graphs
with the higher symmetry $SO(2)\times Z_{2}$, where $Z_{2}$ is the discrete
symmetry
$$ Z_2\;\;:\;\;3\leftrightarrow 4\;,\; 5 \leftrightarrow 6\;.\eqno(11.3.3) $$
The new irreducible graphs which have this symmetry are the 2 self K$_{SO(6)
/SO(2)}$-comple-\break
mentary graphs (and the 2 self K$_{SO(6)+SO(2)}$-conjugate
graphs) in the 2 quartets of Fig. 7 and the $SO(2)$-invariant graph quartet in
Fig. 11.

The $SO(2)\times Z_{2}$-invariant graphs of order 6 are collected in the
7-dimensional subansatz of (11.3.1)
$$ L_{13}=L_{23}=L_{14}=L_{24} $$
$$ L_{15}=L_{25}=L_{16}=L_{26} $$
$$ L_{35}=L_{46}\;,\;L_{45}=L_{36} $$
$$ L_{12}\;,\;L_{34}\;,\;L_{56}\eqno(11.3.4) $$
which is called $SO(6)[d(SO(2)),7']$ in the nomenclature of Ref. [25].
The reduced master equation of $SO(6)[d(SO(2)),7']$
$$\eqalignno{ L_{13}(1-xL_{13})&=L_{13}(2L_{13}+2L_{15}+L_{12}+L_{34}+L_{35}+
L_{45})\cr
&\phantom{=L}-L_{13}(L_{12}+L_{34})-L_{15}(L_{35}+L_{45})\cr
L_{15}(1-xL_{15})&=L_{13}(2L_{13}+2L_{15}+L_{12}+L_{56}+L_{35}+L_{45})\cr
&\phantom{=L}-L_{13}(L_{35}+L_{45})-L_{15}(L_{12}+L_{56})\cr
L_{35}(1-xL_{35})&=L_{35}(2L_{13}+2L_{15}+2L_{45}+L_{34}+L_{56})\cr
&\phantom{=L}-2L_{13}L_{15}-L_{45}(L_{34}+L_{56})\cr
L_{45}(1-xL_{45})&=L_{45}(2L_{13}+2L_{15}+2L_{35}+L_{34}+L_{56})\cr
&\phantom{=L}-2L_{13}L_{15}-L_{35}(L_{34}+L_{56})\cr
L_{12}(1-xL_{12})&=4L_{12}(L_{13}+L_{15})-2(L_{13}^2+L_{15}^2)\cr
L_{34}(1-xL_{34})&=2L_{34}(2L_{13}+L_{35}+L_{45})-2(L_{13}^2+L_{35}L_{45})\cr
L_{56}(1-xL_{56})&=2L_{56}(2L_{15}+L_{35}+L_{45})-2(L_{15}^2+L_{35}L_{45} )&(
11.3.5)\cr}$$
follows as above from Eqs (7.2.4) and (11.3.4).

Except for possible sporadic solutions, we have solved the 7-dimensional
subansatz completely and we present the new solutions in 2 groups.

The $(SO(2)\times Z_{2})$-invariant level-families which form the quartet in
Fig. 11 are
$$L_{13}=\frac{1}{2(x+4)}\left(1-\eta R\left(4-\e(x+4)\sqrt{x(x-4)}\right)
\right) $$
$$L_{15}=\frac{1}{2(x+4)}\left(1-\eta R\left(4+\e(x+4)\sqrt{x(x-4)}\right)
\right) $$
$$ L_{35}=\frac{1}{2(x+4)}\left(1+2\eta(x+2)R\right)+
{\s \eta \over 2} \sqrt{{x^2-4x-4  \over x^4-16x^2+16}} $$
$$ L_{45}=\frac{1}{2(x+4)}\left(1+2\eta(x+2)R\right)-
{\s \eta \over 2} \sqrt{{x^2-4x-4  \over x^4-16x^2+16}} $$
$$L_{12}=\frac{1}{2(x+4)}\left(1+\eta\left(16 R-\xi {x+4 \over x}\right)\right)
$$
$$L_{34}=\frac{1}{2(x+4)}\left(1-\eta R\left(x^2+2x-4+2\e(x+4)\sqrt{(x-4)/x}
\right)\right)
$$
$$L_{56}=\frac{1}{2(x+4)}\left(1-\eta R\left(x^2+2x-4-2\e(x+4)\sqrt{(x-4)/x}
\right)\right)
$$
$$ c=\frac{1}{2(x+4)}\left(15x-\eta(x-2)(x^2+4)R\right)-{(1+\xi)\eta \over 2}
$$
$$ R \equiv (x^4-16x^2+16)^{-1/2}\eqno(11.3.6) $$
where $\eta=\pm 1$ is K-conjugation on $SO(6)$, $\xi=\pm 1$ labels the left and
right sides of the quartet and $\sigma$,$\epsilon=\pm 1$
label 4 automorphic copies of the quartet.
More precisely, the high-level limit of the level-families establishes the
correspondence
$$ SO(6)^{\#}[d(SO(2)),7']_1:\;\;\;\eta=\xi=1 \eqno(11.3.7a)$$
$$ SO(6)^{\#}[d(SO(2)),7']_2:\;\;\;\eta=-\xi=1 \eqno(11.3.7b)$$
with the graphs in Fig. 11, where $\sigma=\epsilon=1$ is required to obtain the
labelling shown.
The rest of the quartet is obtained by K-conjugation of these two
level-families.

The level-families of this quartet have generically irrational central charge,
and they are unitary for integer level $x\geq 5$.
The value at level 5
$$ c\left( SO(6)_5^{\#}[d(SO(2)),7']_1\right)=
\frac{1}{18}\left(57-{87 \over \sqrt{241}}\right)\simeq 2.8553 \eqno(11.3.8)$$
is the lowest unitary irrational central charge in the quartet, and all the
central charges of the quartet increase monotonically with the level toward
the integers $c_{0}=$dim$E(G)$.

\bigskip\noindent
{\bf 11.4.} {\bf The Self K$_{g/h}$-Conjugate Level-Families of
SO(6)$_{diag}$}
\bigskip

The $SO(2)$-invariant quartets of Fig. 7 contain the only 2 self
K$_{g/h}$-complementary graphs of order 6, where $g/h=SO(6)/SO(2)$.
The corresponding level-families of these quartets
$$ L_{13}=\frac{1}{2(x+4)}\left(1+\eta \e\sqrt{{x+4 \over x}}\right) $$
$$ L_{15}=\frac{1}{2(x+4)}\left(1-\eta \e\sqrt{{x+4 \over x}}\right) $$
$$ L_{35}=\frac{1}{2(x+4)}+{\s\eta \over 2\sqrt{x(x+4)}} $$
$$ L_{45}=\frac{1}{2(x+4)}-{\s\eta \over 2\sqrt{x(x+4)}} $$
$$ L_{12}=\frac{1}{2(x+4)}\left(1-\eta{x+4 \over x}\right) $$
$$ L_{34}=\frac{1}{2(x+4)}\left(1-{2\eta\e \over x}\sqrt{{x+4 \over x}}
+{\eta\e\xi \over x}\sqrt{{(x+4)(x^2+4) \over x}}\right)  $$
$$ L_{56}=\frac{1}{2(x+4)}\left(1+{2\eta\e \over x}\sqrt{{x+4 \over x}}
-{\eta\e\xi \over x}\sqrt{{(x+4)(x^2+4) \over x}}\right) $$
$$ c=\frac{15x}{2(x+4)}-\frac{\eta}{2}\eqno(11.4.1) $$
are also solutions of (11.3.5), where $\eta= \pm 1$ is K-conjugation, and
$\sigma$,$\epsilon= \pm 1$ label automorphic copies.
The values $\xi=1$ or $-1$ correspond to solutions in the left and right
quartets of Fig. 7 respectively.
The level-families of both quartets are unitary on all integer levels $x\geq
1$.

More precisely, the high-level limit of the level-families (11.4.1)
establishes the correspondence
$$ SO(6)^{\#}[d(SO(2)),7']_3:\;\;\;\eta=\xi=1\eqno(11.4.2a) $$
$$ SO(6)^{\#}[d(SO(2)),7']_4:\;\;\;\eta=-\xi=1\eqno(11.4.2b) $$
where $3$ and $4$ are the level-families of the left and right self
K$_{SO(6)/SO(2)}$-comple-\hfil
mentary graphs respectively in Fig. 7.
Both of these self K$_{SO(6)/SO(2)}$-conjugate level-families
are ${\widehat SO(2)}$-invariant with $SO(2)=J_{12}$, and, in agreement with
Eq. (10.2.1), they have central charges $c=c_{SO(6)/SO(2)}/2$.
Their K-conjugate level-families, which are self
K$_{SO(6)+SO(2)}$-conjugate, have $c=(c_{SO(6)}+c_{SO(2)})/2$.

It is clear from the square roots in (11.4.1) that these self
K$_{g/h}$-conjugate constructions have generically irrational conformal
weights.
Since this situation was also observed for self K-conjugate
constructions \cite{25,26,33},
we expect that irrational conformal weights will occur generically in all
the generalized self K-conjugate constructions.

\bigskip
\centerline{\bf Acknowledgements}
\bigskip

We thank F. Harary for a helpful discussion.

Two of the authors (M.B.H. and E.K.) are grateful to T. Eguchi, T. Miwa
and the members of RIMS for their kind invitation to participate in the
``Infinite Analysis" workshop, and for their financial support and hospitality.
E.K. would also like to thank Prof. T. Inami for hospitality at the Yukawa
Institute for Fundamental Physics.

\bigskip\noindent
{\large\bf Appendix A.} {\large\bf Counting (0,0) and (1,0) Operators }
\bigskip

To study (0,0) and (1,0) operators of Lie $h$-invariant constructions, recall
the general $T\, J$ algebra$^{15}$
$$ [L^{(m)},J_{a}^{(n)}]=-n{M_{a}}^{b}(L)J_{b}^{(m+n)}+{N_{a}}^{bc}(L)
T_{bc}^{(m+n)}\eqno(A.1a)$$
$$  {M_a}^b(L)=2G_{ac}L^{cb}+{f_{ad}}^{e}L^{dc}{f_{ce}}^b  \eqno(A.1b)$$
$$ {N_a}^{bc}(L)=-i {f_{ad}}^{(b}L^{c)d}\eqno(A.1c)$$
of any conformal construction $L^{ab}$ on affine $g$.
The matrix ${M_{a}}^{b}(L)$ controls the spectrum of the level-one current
states in the affine vacuum module,
$$L^{(m)}\psi^{a}_{i}J_{a}^{(-1)}|0\rangle =\delta_{m,0}\Delta_{i}\,
\psi^{a}_{i}J_{a}^{(-1)}|0\rangle\,\,\,,\,\,\,m\geq 0\eqno(A.2a)$$
$$\psi_{i}^{b}{M_{a}}^{b}(L)=\Delta_{i}\,\psi_{i}^{a}\,\,\,,\,\,\,i=1,\cdots,
{\rm dim}\,g\eqno(A.2b)$$
where $\psi_{i}^{a}$ are the left eigenvectors of ${M_{a}}^{b}(L)$ and
$\Delta_{i}$ are the corresponding conformal weights.
It is also known$^{15}$ that $M_{ab}(L)={M_{a}}^{c}(L)G_{cb}$ is a symmetric
matrix.

Consequences of unitarity on integer levels of affine compact $g$ are obtained
as follows.
Work for simplicity in any Cartesian frame of $g$,
where ${f_{ab}}^{c}$, $G_{ab}$ and all unitary constructions $L^{ab}$ are real.
It follows that ${M_{a}}^{b}(L)$, $M_{ab}(L)$ and $\Delta_{i}$ are real, and we
may choose orthonormality in the form
$$\psi_{i}^{a\,*}G_{ab}\psi_{j}^{b}=\delta_{ij}\,.\eqno(A.3)$$
The stronger result
$$0\leq \Delta_{i}\leq 1\eqno(A.4)$$
follows by the standard operator unitarity argument on K-conjugate partners,
which satisfy $\Delta(L)+\Delta({\tilde L})=\Delta(L_{g})$ for all conformal
weights.

Restrict attention now to any conformal construction whose Lie symmetry is
exactly $h$.
In this case, the construction also satisfies Eq. (4.1)
$$  \d L^{ab}(\psi_{h})=-i\psi_{h}^{c}\,{N_c}^{ab}(L)   =0 \eqno(A.5)$$
where $\psi_{h}$ parametrizes $H$ in the vicinity of the origin,
so it follows from (A.1) that
$$[L^{(m)},\psi_{h}^{a}J_{a}^{(n)}]=-n\psi_{h}^{a}{M_{a}}^{b}(L)
J_{b}^{(m+n)}\,.\eqno(A.6)$$
Next, consider the state
$$L^{(-1)}\psi_{h}^{a}J_{a}^{(-1)}|0\rangle =\psi_{h}^{a}{M_{a}}^{b}(L)J_{b}^{(
-2)}|0\rangle\eqno(A.7)$$
in a unitary Lie $h$-invariant theory and compute its norm from each of
the two forms (A.7) to obtain
$$\psi^{d}_{h}{M_{d}}^{b}(L)G_{ba}(\psi_{h}^{a}-\psi_{h}^{c}{M_{c}}^{a}
(L))^{*}=0\,\,.\eqno(A.8)$$
A more transparent form of this identity
$$\sum_{i=1}^{{\rm dim}\,g}\Delta_{i}(1-\Delta_{i})|\psi_{i}^{*\,
b}G_{ba}\psi_{h}^{a}|^{2}=0\eqno(A.9)$$
is obtained with (A.3) by expanding $\psi_{h}^{a}=\sum_{i}c_{i}(\psi_{h})
\psi_{i}^{a}$ in the left eigenbasis of ${M_{a}}^{b}(L)$.
It follows from (A.9) that
$$[L^{(m)},\psi_{i}^{a}J_{a}^{(n)}]=-n\Delta_{i}\psi^{a}_{i}J_{a}^{(m+n)}
\eqno(A.10a)$$
$$\Delta_{i}=0\,\,{\rm or}\,\,1\,\,\,{\rm when}\,\,\,
\psi_{i}^{*\,a}G_{ab}\psi_{h}^{b}\not= 0\eqno(A.10b)$$
so all the currents of $h$ are (0,0) or (1,0) operators of
a unitary Lie $h$-invariant construction.

The converse of this result was obtained in Ref. 33: When $L$ is unitary  and
$\psi^{a}$ is a left eigenvector of ${M_{a}}^{b}(L)$
with $\Delta_{i}=0$ or $1$, then  $\d L^{ab}(\psi)=-i\psi^{c}{N_c}^{ab}(L)=0$.
It follows that the set of all (0,0) and (1,0) currents of a unitary
theory generate a Lie symmetry, so that a unitary theory with a Lie
symmetry which is exactly $h$ contains
$${\rm dim}\,h=N_{0}+N_{1}\eqno(A.11)$$
(0,0) and (1,0) currents, where $N_{0}$ and $N_{1}$ are the numbers of each
type respectively.
This completes the proof of the theorem in Section 5.\hfill$\qed$

We can also show that
$$h=h_{0}\oplus h_{1}\eqno(A.12)$$
where $h_{0}$ and $h_{1}$ are the commuting affine subalgebras generated by the
(0,0) currents $J_{A}$ and the (1,0) currents $J_{I}$ respectively of $L^{ab}$,
$$[L^{(m)},J_{A}^{(n)}]=0\,\,\,,\,\,\,[L^{(m)},J_{I}^{(n)}]=-nJ_{I}^{(m+n)}\,.
\eqno(A.13)$$
To see this, note first that the commutator of two (0,0) currents commutes
with the stress tensor, so that the set of (0,0) currents of $L^{ab}$ is
closed under commutation.
Similarly, the set of (1,0) currents of $L^{ab}$ is closed under
commutation because these currents
and their commutators are (0,0) currents of the K-conjugate construction
${\tilde L}^{ab}$, and hence (1,0) currents of $L^{ab}$.
Finally, the Jacobi identity and current algebra
$$\left[ L^{(m)},[J_{A}^{(p)},J_{I}^{(q)}]\right]=-q[J_{A}^{(p)},J_{I}^{(m+q)}]
\eqno(A.14a)$$
$$[J_{A}^{(p)},J_{I}^{(q)}]=i{F_{AI}}^{a}J_{a}^{(p+q)}+C_{AI}p\delta_{p+q,
0}\eqno(A.14b)$$
imply that ${F_{AI}}^{a}$ and $C_{AI}$ are zero, so that the (0,0) currents
commute with the (1,0) currents.
This completes the proof of the proposition in Section 5.\hfill$\qed$

\end{document}